\documentclass[journal,twoside,letterpaper]{IEEEtran}
\usepackage{amsmath,amsfonts}
\usepackage{algorithm2e}
\usepackage{array}
\usepackage{url}
\usepackage{cite}
\usepackage{url}
\usepackage{booktabs}
\usepackage{braket} 

\usepackage{subcaption}

\usepackage{mathtools}
\DeclarePairedDelimiter{\abs}{\lvert}{\rvert}

\newcommand{\mytitle}{From Graphs to Qubits: A Critical Review of Quantum Graph Neural Networks}

\begin{document}

\title{\mytitle}

\author{Andrea~Ceschini,~\IEEEmembership{Student Member,~IEEE,}
Francesco~Mauro,~\IEEEmembership{Student Member,~IEEE,} 
Francesca~De~Falco,~\IEEEmembership{Student Member,~IEEE,} 
Alessandro~Sebastianelli, 
Alessio~Verdone,~\IEEEmembership{Student Member,~IEEE,}
Antonello~Rosato,~\IEEEmembership{Member,~IEEE,}
Bertrand~Le~Saux, 
Massimo~Panella,~\IEEEmembership{Senior Member,~IEEE,}
Paolo~Gamba,~\IEEEmembership{Fellow,~IEEE,}
and~Silvia~L.~Ullo,~\IEEEmembership{Senior Member,~IEEE}
\thanks{A. Ceschini, F. De Falco, A. Verdone, A. Rosato and M. Panella are with the Department of Information Engineering, Electronics and Telecommunications, University of Rome ``La Sapienza'', 00184 Rome, Italy (e-mail: andrea.ceschini@uniroma1.it, massimo.panella@uniroma1.it).}
\thanks{F. Mauro and S.L. Ullo are with the Department of Engineering, University of Sannio, 82100 Benevento, Italy (email: f.mauro@studenti.unisannio.it, ullo@unisannio.it).}
\thanks{A. Sebastianelli and B. Le Saux are with the European Space Agency, 00044 Frascati, Italy (email: Alessandro.Sebastianelli@esa.int)}
\thanks{P. Gamba is with the Department of Electrical, Computer and Biomedical Engineering, University of Pavia, 27100 Pavia, Italy (e-mail:  paolo.gamba@universitadipavia.it).}
\thanks{Corresponding author: Prof. M. Panella, Via Eudossiana 18, 00184 Rome, Italy (phone: +39-0644585496; e-mail: massimo.panella@uniroma1.it).}
\thanks{Manuscript received August, 2024.}}
\markboth{S\MakeLowercase{ubmitted to} IEEE Transactions on Pattern Analysis and Machine Intelligence}{A. Ceschini, \MakeLowercase{\textit{(et al.)}}: \uppercase{\mytitle}}

\maketitle

\begin{abstract}
Quantum Graph Neural Networks (QGNNs) represent a novel fusion of quantum computing and Graph Neural Networks (GNNs), aimed at overcoming the computational and scalability challenges inherent in classical GNNs that are powerful tools for analyzing data with complex relational structures but suffer from limitations such as high computational complexity and over-smoothing in large-scale applications. Quantum computing, leveraging principles like superposition and entanglement, offers a pathway to enhanced computational capabilities. This paper critically reviews the state-of-the-art in QGNNs, exploring various architectures. We discuss their applications across diverse fields such as high-energy physics, molecular chemistry, finance and earth sciences, highlighting the potential for quantum advantage. Additionally, we address the significant challenges faced by QGNNs, including noise, decoherence, and scalability issues, proposing potential strategies to mitigate these problems. This comprehensive review aims to provide a foundational understanding of QGNNs, fostering further research and development in this promising interdisciplinary field.
\end{abstract}


\section{Introduction}
\IEEEPARstart{G}{raph} Neural Networks (GNNs) have emerged as powerful tools for analyzing and processing data represented in graph structures, which is a format ubiquitous in diverse fields such as social networks, recommendation systems, cybersecurity, sensor networks, and natural language processing. Since their introduction \cite{scarselli2008graph,gori2005new,kipf2016semi,Bruna19,hamilton2017inductive,DuvenaudMABHAA15,GNNGama,gilmer2017neural}, GNNs have demonstrated remarkable effectiveness in various graph-related tasks thanks to their ability to learn and generalize from complex structural relationships encoded within nodes and edges. At their core, GNNs operate by iteratively aggregating information from neighboring nodes to learn representations for each node, as determined by the graph's topology. However, one significant limitation of GNNs is their computational complexity. The iterative nature of the message-passing operations requires substantial computational resources, especially for large-scale graphs with millions of nodes and edges \cite{abadal2021computing}. This often results in high memory usage and prolonged training times, posing challenges for scalability and efficiency. Moreover, GNNs tend to suffer from over-smoothing, where repeated aggregation of node features causes the node representations to become indistinguishable from one another, thereby reducing their discriminative power \cite{zhang2023comprehensive}.

Researchers have begun exploring quantum computing approaches to overcome some of these limitations and enhance the computational capabilities of neural models. Quantum Machine Learning (QML) models offer provable advantages over classical counterparts in regression and classification tasks, applicable to both classical and quantum data settings \cite{9647979,abbas2021power,liu2021rigorous,caro2022generalization}. Additionally, quantum computers have shown exponential advantages in analyzing data from quantum processes \cite{huang2022advantage}. Quantum Neural Networks (QNNs) have emerged as a promising trend in Deep Learning (DL), leveraging quantum computing principles such as superposition and entanglement to handle large-scale heterogeneous data more efficiently than classical neural networks (NNs) \cite{tacchino2019artificial,tian2023recent}. 

However, adopting a quantum framework presents challenges, including system coherence and qubits connectivity. Despite the current technical infeasibility of fault-tolerant quantum computers, progress in Noisy Intermediate-Scale Quantum (NISQ) devices has been significant, enabling Variational Quantum Circuits (VQCs) as a pragmatic approach to harness quantum resources \cite{preskill2018quantum,mitarai2018quantum,tacchino2020variational,cerezo2021variational}. QNNs, particularly implemented via Parameterized Quantum Circuits (PQCs), have shown suitability for NISQ devices, with VQCs enabling their execution on current quantum hardware, impacting real-world applications \cite{abbas2021power,mangini2022quantum}. However, QNNs lack the non-linear complexities of classical DL models, and they face challenges such as barren plateaus, where gradient variance exponentially decays, hindering training \cite{havlivcek2019supervised,holmes_2021_nonlinear,cerezo2021cost}.

In order to address these drawbacks and enhance the applicability of GNNs to complex systems, recent advancements have led to the development of Quantum Graph Neural Networks (QGNNs). These novel frameworks integrate principles from quantum computing with the versatility of GNNs, aiming to better capture and utilize the rich hierarchical and relational information present in graph-structured data.
In this review, we explore the foundational concepts, methodologies, and emerging applications of QGNNs, highlighting their potential to revolutionize the analysis and understanding of complex interconnected systems.

\begin{figure*}[!ht]
    \centering
    \includegraphics[width=\textwidth]{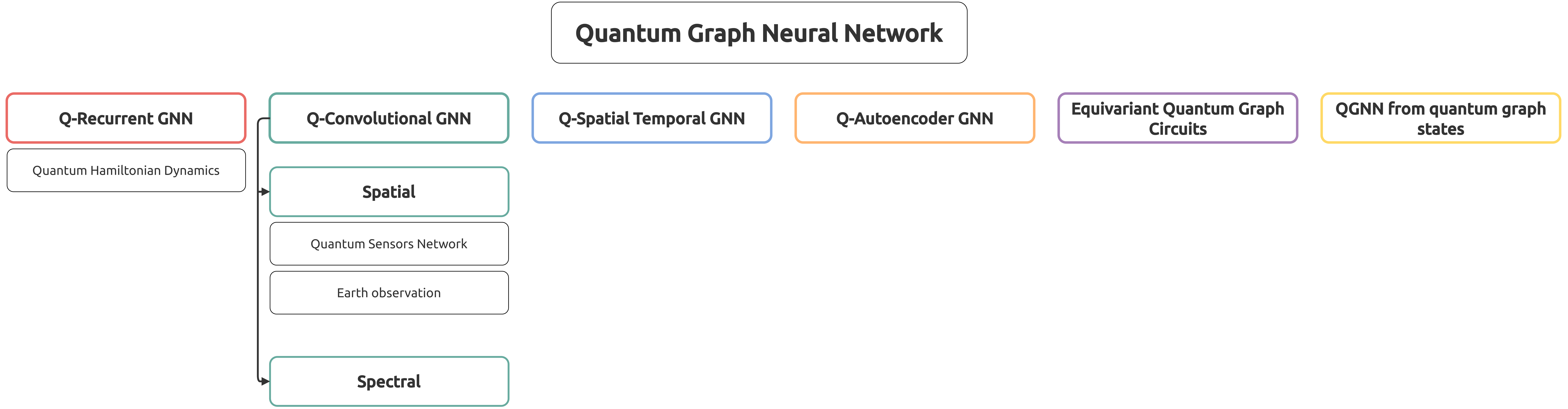}
    \caption{Main scheme of the review.}
    \label{fig:schema_review_main}
\end{figure*}

This article makes several notable contributions, summarized as follows:
\begin{enumerate}
    \item Comprehensive Review: We present the most comprehensive survey of modern QGNNs to date. This includes in-depth descriptions of key models, critical comparisons between various approaches, and a synthesis of the underlying algorithms. Our review serves as a crucial resource for researchers seeking a thorough understanding of the current landscape of QGNNs.

    \item Extensive Resource Compilation: We gather and organize an extensive collection of resources on QGNNs, including state-of-the-art models, benchmark datasets, open-source codebases, and notable practical applications in fields such as quantum chemistry and materials science. In this review, the selection of papers was conducted using two primary platforms: IEEE Xplore and Google Scholar. Our approach involved identifying both the most recent and the most frequently cited papers within the relevant fields. This article serves as a practical guide for both understanding and utilizing different quantum deep learning approaches across diverse real-life applications.
    
    \item Architecture Analysis: We provide a thorough analysis of the theoretical aspects of QGNNs, highlighting the limitations of current methods such as scalability issues and noise susceptibility. Based on this analysis, we suggest future research directions, including the development of more efficient quantum algorithms, noise-resilient architectures, and improved quantum hardware integration to enhance the practicality and performance of QGNNs.

    \item Taxonomy and Classification: For the first time in literature, we introduce a comprehensive taxonomy of QGNNs as shown in Fig.~\ref{fig:schema_review_main}, categorizing existing approaches based on their architectural designs, computational strategies, and application domains. This classification framework aids in systematically understanding the diverse methodologies and their respective strengths and weaknesses.

    \item Practical Guidance: We provide practical insights for implementing QGNNs, addressing key questions such as selecting the most appropriate QGNN variant or ansatz for specific problem types and optimizing their performance effectively. This section aims to bridge the gap between theoretical research and practical deployment, ensuring that our findings are accessible and applicable to a broad audience.
\end{enumerate}
With these contributions, our article not only advances the understanding of QGNNs but also provides practical tools and insights for their application and further development, ensuring a robust foundation for future research and innovation in the field.

The structure of this paper is outlined as follows. In Sect.~\ref{sec:gnn}, we present an overview of the fundamental concepts behind classical GNNs, along with illustrative application examples. Sect.~\ref{sec:fundamentals} covers the foundational principles of quantum computing and the functioning of VQCs. Sect.~\ref{sec:qgnns} offers an in-depth analysis of Quantum Graph Neural Networks (QGNNs), discussing various components such as the ansatz, computational efficiency, solution quality, the impact of noise and errors, and hardware-specific implementations. In Sect.~\ref{sec:applications}, we explore the application of QGNN algorithms and assess their potential for achieving quantum advantage. Sect.~\ref{sec:challenges} addresses the challenges and limitations associated with current QGNN implementations on NISQ devices. 
Finally, in Sect.~\ref{sec:conclusions} we conclude the paper with a summary of key findings and their implications for future research in this field.

\section{Preliminaries on Graph Neural Networks} \label{sec:gnn}

We present the basic concepts of standard GNNs since the QGNNs are built upon these foundational ideas, introducing quantum mechanics principles to further enrich the model's expressiveness and computational power.

\subsection{Definitions and Notations}
Graphs are mathematical structures that represent a set of objects (nodes) and the pairwise relations (edges) existing between them. Many real-life networks can be described as graphs such as transportation, social telecommunication or energy ones, chemical molecules, multi-agent systems or atmospheric phenomena.
Formally, a graph $\mathcal{G}$ can be defined as $\mathcal{G} = (\mathcal{N}, \mathcal{E})$, where $\mathcal{N}$ and $\mathcal{E}$ are the set of nodes and edges respectively. Graphs with directional edges are directed graphs, conversely, they are undirected. 


A graph with $n$ nodes can be represented in matrix form through the Adjacency matrix $\mathbf{A} \in \mathcal{R}^{n \times n}$, where $A_{ij} = 1$ if $n_i$ and $n_j$ are connected, else $0$.
The Degree matrix $\mathbf{D}$ is a diagonal matrix which represents the number of edges connected to each node $n$, where $D_{ii} = \sum_{j \in V} A_{ij}$.
Laplacian Matrix $\mathbf{L}$ encodes the whole structure of the graph in matrix form: $\mathbf{L} = \mathbf{D} - \mathbf{A}$. Imbalanced weights may undesirably affect the matrix spectrum, leading to the need of normalization: $\mathbf{\tilde{A}} = \mathbf{A}+\mathbf{I}$, $\mathbf{\tilde{D}}$ is built from $\mathbf{\tilde{A}}$ and finally ${\mathbf{L}_\mathrm{norm} = \mathbf{{D}}^{-1/2}\mathbf{{L}}\mathbf{{D}}^{-1/2} = \mathbf{I} - \mathbf{{D}}^{-1/2}\mathbf{{A}}\mathbf{{D}}^{-1/2}}$.

A feature vector of arbitrary dimensions $f$ can be linked to each node (or also edge): the node feature matrix $\mathbf{X} \in \mathcal{R}^{n \times f}$ represent all the features of the graph, where $X_{ij}$ represents the $j$-th feature of the $i$-th node.

Analyzing graph data with traditional Machine Learning (ML) algorithms poses several challenges \cite{chami2022machine}: $1)$ conventional ML and DL tools are specialized in structured data types, such as images or audio data, on which different samples share the same structure and size, grid and sequences, but differ in values; $2)$ on the contrary, graphs are unstructured data types, they don't have a fixed form or a constant size and also differ in the number of nodes and the relationships between them.

\subsection{Architecture of Graph Neural Networks}
\begin{figure}[!ht]
    \centering
    \includegraphics[width=0.79\columnwidth]{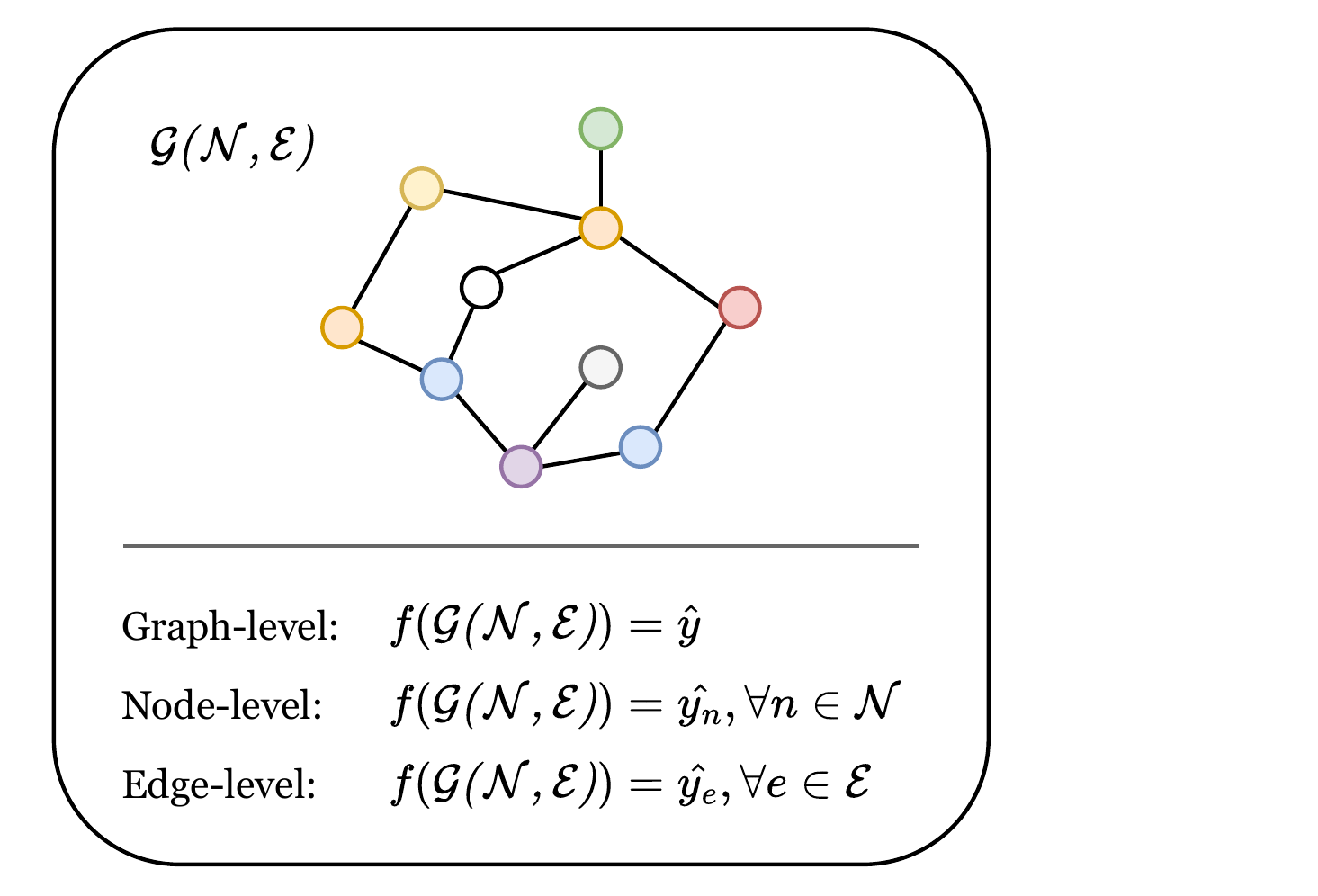}       
    \caption{Illustration of node-classification and graph-classification tasks}
    \label{fig:graph_prediction_types}
\end{figure}

GNNs have been developed to works in an efficient way with graph data: they are specialized for tasks such as node-level, edge-level, and graph-level regression or classification, as illustrated in Fig.~\ref{fig:graph_prediction_types}.

The more important mechanism at the core of GNNs is the Graph Convolutional Layers (GCL); in sequence more GCLs form a GNN. A GCL relies on 3 main properties:
\begin{itemize}
    \item \textbf{Locality}: it leverages the inherent locality of graph structures to perform convolutions on nodes and their neighborhoods.
    \item \textbf{Permutation Equivariant}: to not depend on the arbitrary ordering of the features of the graph when inserted in matrix form for enabling differentiable mechanism.
    \item \textbf{Differentiable}: it defines a trainable matrix of weights $W \in \mathcal{R}^{f \times f'}$ that operate together to the message passing-based stream of information between neighboring nodes, defined by the adjacency matrix $A$.
\end{itemize}

Mathematically, a GCL can be represented by
\begin{equation}
\label{eq:gcl}
    f(\mathbf{X},\mathbf{A}) = \varphi(\mathbf{A}(\mathbf{X}\mathbf{W}))\,,
\end{equation} 
where $\varphi$ is a differentiable activation function.
%
Stacked one upon the other, GCLs allow iteratively to update node or edge representations, incorporating information from neighboring nodes and allowing it to flow through the graph structure: by definition, by stacking n GCLs, the information flow concerns the $n$-th hop neighbor.

Relying on this formalism, the research community has produced several types of GNNs, which can be mainly categorized into Recurrent Graph Neural Networks (RecGNNs), Convolutional GNNs (ConvGNNs), Graph Autoencoders (GAEs), Graph Transformer (GT), and Spatial–Temporal GNNs (STGNNs).

Each type offers unique advantages and challenges. RecGNNs, pioneered by \cite{sperduti1997supervised,micheli2004contextual}, focus on information diffusion through iterative updates, while ConvGNNs, introduced by \cite{kipf2016semi}, extending the concept of convolutional neural networks to graph data, enabling effective information propagation among neighboring nodes. 
Graph autoencoders, such as those developed by \cite{kipf2016variational}, aim to learn low-dimensional representations of graphs in an unsupervised manner. They consist of an encoder that maps the input graph to a latent space and a decoder that attempts to reconstruct the graph from this latent representation. 

More recently, Spatio-temporal GNNs (STGNNs), introduced by \cite{yu2017spatio}, are designed to handle data that evolves over space and time. These models combine graph convolutional layers with temporal processing units, such as recurrent neural networks or attention mechanisms, to capture the dynamic nature of the data. 
Graph Transformers \cite{graph_transformer} merges graph learning or transformer architecture demonstrating powerful versatility and performances across different tasks. 
These advancements in GNNs offer powerful tools for analyzing and understanding complex graph-structured data, opening avenues for various applications across domains.

\subsection{Applications}
From a mathematical point of view, GNNs address to solve tasks at the node, edge and graph level. Classification and regression tasks can be applied at all three levels: for example, node regression involves determining continuous values of individual nodes \cite{ma2021regression} or graph classification focuses on categorizing entire graphs, similar to how images are classified in image recognition tasks, to a precise label \cite{quek2011structural}. Link predictions have gained particular popularity in social scenarios \cite{daud2020applications}: predicting the existence of a link between two nodes can be useful in suggesting new friendships in a social network. Other important GNN tasks are graph similarity (or graph matching), which aims to evaluate the similarity between two graphs, Graph Generation or Conditioned Graph Generation, very important in chemistry domains.


GNNs have proven to be highly effective, impacting a diverse array of fields.
In computer vision, GNNs are used to process images that can be represented as graphs with pixels as nodes and edges connecting adjacent pixels. This approach enables tasks like scene graph generation, where images are parsed into semantic graphs depicting objects and their relationships \cite{pradhyumna2021graph}.

Natural Language Processing (NLP) employs GNNs in several tasks: they can be used to represent texts and the complex relationships among words as graphs. GNNs are applied to these structures to perform tasks such as text classification and machine translation \cite{wu2023graph} or also modern applications in combination with Large Language Model (LLM) and Retrieval Augmented Generation (RAG) paradigms \cite{jaiswal2024all, li2024graphneuralnetworkenhanced}.

Traffic forecasting benefits from STGNNs, which model traffic networks by merging temporal and spatial features of the flow of traffic and the road network respectively. Nodes can represent crossroads while edges are roads, and the GNN has to predict future time speed on the network \cite{jiang2022graph, shin2024pgcn, ju2024cool}.

In chemistry, GNNs are used to explore molecular structures, where atoms are nodes and chemical bonds are edges. This application aids in understanding molecular properties and interactions \cite{reiser2022graph},also by incorporating explainability insights \cite{wu2023chemistry}.
Earth observation (EO) is another emerging application of GNNs with several challenges to face: earthquake detection \cite{bilal2022early}, complex atmospherically phenomena \cite{mauro2024hybrid} or solar \cite{verdone2024explainable} and wind \cite{liu2023spatio } power forecasting  are recent field of applications of GNNs  where capturing distant and non-local dependencies is crucial.

Other important fields where the GNNs have been applied with success are social networks \cite{awasthi2023gnn}, anomaly detection \cite{tang2024gadbench}, and recommender systems \cite{gao2023survey}.


\section{Fundamentals of Quantum Computing}\label{sec:fundamentals}
\subsection{Quantum States}
In contrast to classical computers, which operate with bits, the fundamental building block of a quantum computer is the qubit. The latter encompasses two fundamental states: $\ket{0} = \begin{pmatrix} 1 \\ 0 \end{pmatrix}$ and $\ket{1} = \begin{pmatrix} 0 \\ 1 \end{pmatrix}$, representing the ground and excited states within a two-tier quantum environment. Any pure quantum state within the same Hilbert space can be expressed as $\ket{\psi} = \alpha\ket{0} + \beta\ket{1}$, where the probability amplitudes $\alpha, \beta \in \mathbb{C}$ are complex numbers fulfilling $|\alpha|^2 + |\beta|^2 = 1$, or $\ket{\psi} = e^{i\gamma} \left(\cos(\frac{\theta}{2})\ket{0} + e^{i\phi} \sin(\frac{\theta}{2})\ket{1} \right)$, where $\theta$, $\phi$, and $\gamma$ are real numbers with $0 \leq \theta \leq \pi$ and $0 \leq \phi < 2\pi$.

A quantum register, comprising an $n$-qubit space, arises from the tensor product of $n$ single-qubit Hilbert spaces:
\begin{equation}\begin{split}
    \ket{\psi} & = \alpha_{00...0}\ket{00...0} + \alpha_{00...1}\ket{00...1} + \ldots + \alpha_{11...1}\ket{11...1}\\ 
    & = \sum_{x \in \{0,1\}^n} \alpha_x\ket{x}\,.
\end{split}\end{equation}
This involves a summation across all possible bit strings of length $n$, where the coefficients $\alpha_x$, with $x \in \{0,1\}^n$, are complex numbers linked to each basis state. The tensor product $a \otimes b$ is a mathematical operation used to combine states of two different quantum systems $a$ and $b$. For a pair of qubits, $\ket{\psi_1}$ and $\ket{\psi_2}$, the combined state is depicted as:
\begin{equation}
    \ket{\psi} = \ket{\psi_1}\otimes \ket{\psi_2} = \ket{\psi_1\psi_2}\,.
\end{equation}

This compact notation enables the representation of the state state of a large quantum register without explicitly enumerating all the basis states. Entanglement arises when the quantum state of each particle cannot be described independently of the quantum state of the other particle(s), i.e. the quantum state of the system cannot be decomposed into a tensor product of individual states:
\begin{equation}
    \ket{\psi} \neq \ket{\psi_1}\otimes \ket{\psi_2}\,.
\end{equation}

For instance, the state $\ket{\psi} = \frac{1}{\sqrt{2}}\ket{00} + \frac{1}{\sqrt{2}}\ket{11}$ cannot be decomposed into a tensor product of individual single-qubit states. In general, for a vector $x \in \mathbb{R}^{2^n}$, the encoded amplitude state $\ket{x}$ is defined as:
\begin{equation}
\label{eq:amplitude_encoding}
    \frac{1}{\lVert x \rVert} \sum_{i=1}^{2^n} x_i \ket{i}\,.
\end{equation}

\subsection{Quantum Gates and Quantum Circuits}
\begin{figure}[!ht]
    \centering
    \includegraphics[width=.7\columnwidth]{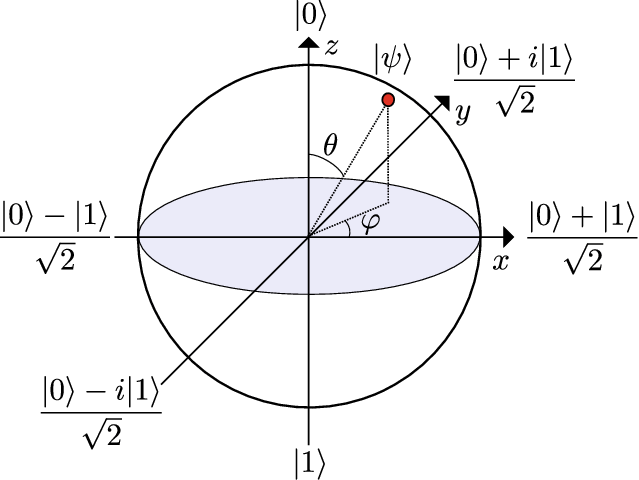}
    \caption{Illustration of qubit states on the Bloch Sphere.}
    \label{fig:bloch_sphere}
\end{figure}

In quantum circuits, all the operations acting on qubits and changing their state are unitary matrices and can be referred to as quantum gates. Single-qubit gates are fundamental operations that manipulate the state of a single qubit on the Bloch Sphere, as illustrated in Fig.~\ref{fig:bloch_sphere}. 

They are detailed in Table~\ref{tab:single_gates} and include the Pauli-$X$, Pauli-$Y$, and Pauli-$Z$ gates, rotating around the $X$, $Y$, and $Z$ axes, respectively. Additionally, Pauli rotation gates $R_x$, $R_y$, and $R_z$ are derived from exponential of the Pauli Hermitian operators, enabling rotation of the state vector by an arbitrary angle $\theta$ around the corresponding Bloch Sphere axis. The Hadamard gate is another fundamental single-qubit gate, inducing superposition states.
\begin{table*}[!ht]
 \caption{Single-qubit Gates}
 \vspace{-3pt}
		\renewcommand\arraystretch{1.2} 
		\setlength{\tabcolsep}{6pt}
    \centering
		\resizebox{0.65\textwidth}{!}{%
    \begin{tabular}{lll}
        \toprule
        Gate & Matrix Representation & Operation \\
				\midrule
        Pauli-X (X-gate) &
        $\begin{bmatrix} 0 & 1 \\ 1 & 0 \end{bmatrix}$ &
        Bit-flip \\
				\rule[-4mm]{-2pt}{1.05cm}
        Pauli-Y (Y-gate) &
        $\begin{bmatrix} 0 & -i \\ i & 0 \end{bmatrix}$ &
        Bit-flip and phase-flip \\
				\rule[-4mm]{-2pt}{1.05cm}
        Pauli-Z (Z-gate) &
        $\begin{bmatrix} 1 & 0 \\ 0 & -1 \end{bmatrix}$ &
        Phase-flip \\
				\rule[-4mm]{-2pt}{1.05cm}
        Hadamard (H-gate) &
        $\frac{1}{\sqrt{2}}\begin{bmatrix} 1 & 1 \\ 1 & -1 \end{bmatrix}$ &
        Superposition \\
				\rule[-4mm]{-2pt}{1.05cm}
        X-Rotation $(R_x)$ &
        $\begin{bmatrix} \cos(\theta/2) & -i\sin(\theta/2) \\ -i\sin(\theta/2) & \cos(\theta/2) \end{bmatrix}$ &
        Rotation about X-axis by angle $\theta$ \\
				\rule[-4mm]{-2pt}{1.05cm}
        Y-Rotation $(R_y)$ &
        $\begin{bmatrix} \cos(\theta/2) & -\sin(\theta/2) \\ \sin(\theta/2) & \cos(\theta/2) \end{bmatrix}$ &
        Rotation about Y-axis by angle $\theta$ \\
				\rule[-4mm]{-2pt}{1.05cm}
        Z-Rotation $(R_z)$ &
        $\begin{bmatrix} e^{-i\theta/2} & 0 \\ 0 & e^{i\theta/2} \end{bmatrix}$ &
        Rotation about Z-axis by angle $\theta$ \\
        \bottomrule
    \end{tabular}}
		\label{tab:single_gates}
\end{table*}

\begin{table*}[!ht]
    \caption{Two-qubit gates}
    \vspace{-3pt}
		\renewcommand\arraystretch{1.2} 
		\setlength{\tabcolsep}{6pt}
    \centering
		\resizebox{0.64\textwidth}{!}{%
    \begin{tabular}{lll}
        \toprule
        Gate & Matrix Representation & Operation \\
				\midrule
        CNOT (Controlled-X gate) &
        $\begin{bmatrix} 1 & 0 & 0 & 0 \\ 0 & 1 & 0 & 0 \\ 0 & 0 & 0 & 1 \\ 0 & 0 & 1 & 0 \end{bmatrix}$ &
        Entangles if control qubit is $|1\rangle$ \\
				\rule[-4mm]{-2pt}{1.35cm}
        SWAP &
        $\begin{bmatrix} 1 & 0 & 0 & 0 \\ 0 & 0 & 1 & 0 \\ 0 & 1 & 0 & 0 \\ 0 & 0 & 0 & 1 \end{bmatrix}$ &
        Swaps states of two qubits \\
        \bottomrule
    \end{tabular}}
		\label{tab:two_gates}
\end{table*}

Two-qubit gates are pivotal for entanglement and interaction among qubits. Examples include the Controlled-NOT (CNOT) gate, which entangles two qubits based on the state of a control qubit, and the SWAP gate, which exchanges the states of two qubits, as shown in Table~\ref{tab:two_gates}. When combined, these gates form a universal set capable of approximating any quantum operation with arbitrary precision. Quantum circuits, composed of sequences of these gates, facilitate the execution of quantum algorithms.

\subsection{Quantum Measurement}
After the data undergoes the unitary transformation of the quantum gate, the transformation result is not directly accessible, and we need to perform quantum measurements to obtain the results. When measured, the qubit collapses to the new state $\ket{0}$ or $\ket{1}$. For example, performing a projective measurement with Pauli-$Z$ observable on the qubit with the state $\ket{\psi} = \alpha\ket{0} + \beta\ket{1}$ generates $1$ (eigenvalue correspondent to eigenvector $\ket{0}$) or $-1$ (eigenvalue correspondent to eigenvector $\ket{1}$) with probability $p(1) = |\alpha|^2$ and $p(-1) = |\beta|^2$, respectively. 

Computational basis measurement can be understood as extracting a sample of a binary string from the distribution defined by the quantum state \cite{nielsen2010quantum}. However, since the measurement results are probabilistic, only one possible value can be obtained in each measurement under some probability; in order to obtain as accurate information about the quantum state as possible, we need to perform repeated measurements.

\subsection{Variational Quantum Algorithms}
\begin{figure*}[!ht]
    \centering
    \includegraphics[width=0.6\textwidth]{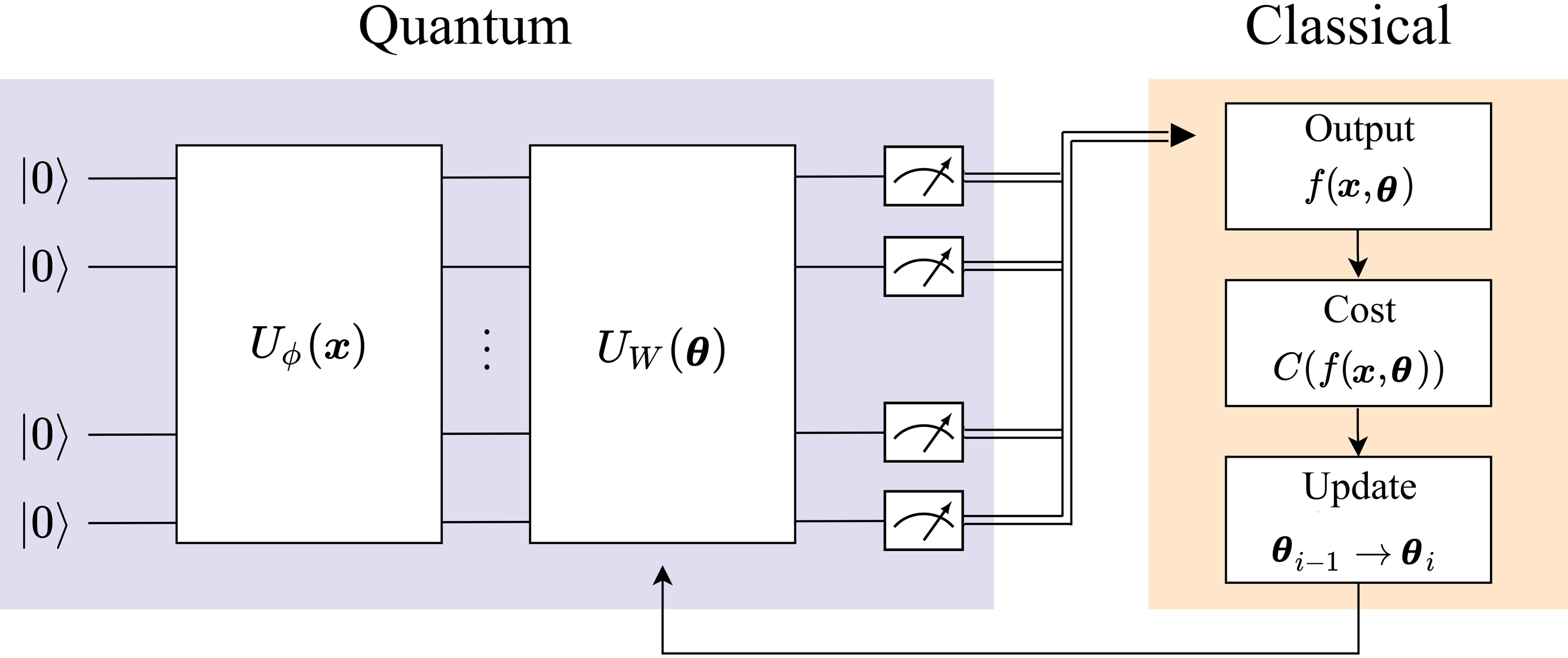}
    \caption{Scheme of a hybrid quantum-classical VQC for supervised learning.}
    \label{fig:VQC_framework}
 \end{figure*}

Variational Quantum Algorithms (VQAs) are based on VQCs and represent the most promising approach to implement QNNs on NISQ devices. A VQA consists of a data encoding stage $U_\phi(\mathbf{x})$, a properly designed ansatz $U_W(\boldsymbol{\theta})$ applied to the resulting quantum state, and a measurement operation at the end of the circuit, as illustrated in Fig.~\ref{fig:VQC_framework}. To encode classical data ${\mathbf{x} \in \mathrm{R}^n}$ into an $n$ qubit quantum circuit, a quantum feature map $\phi: \mathrm{R}^n \rightarrow H^{2^n}$ is applied, where $H^{2^n}$ is a $2^n$ dimensional Hilbert space. It corresponds to applying a unitary matrix $U_{\phi}(\mathbf{x})$ to the initial state $\ket{0}^{\otimes n}$:
\begin{equation}		
	U_{\phi}(\mathbf{x})\ket{0}^{\otimes n} = \ket{\phi(\mathbf{x})} = \ket{\psi}\,. 
\end{equation}

A suitable data encoding strategy is crucial for the attainment of a quantum advantage. An efficient representation of classical input data through quantum states is a necessary prerequisite to benefit from quantum technologies \cite{weigold2021expanding}, and the selection of the unitary $U_{\phi}(\mathbf{x})$ has a considerable impact on the performance of the underlying VQC. Although various data encoding techniques have been proposed in the literature \cite{havlicekSupervisedLearningQuantumenhanced2019,schuld2021machine}, angle encoding is the most popular and efficient method to encode continuous variables into quantum states \cite{abbas2021power,zhaoQDNNDeepNeural2021,weigold2021expanding}. It employs parametrized rotation quantum gates to encode each input feature into a qubit: angle encoding allows to represent $n$ input features by means of $n$ qubits. 

The value of the rotation parameters directly corresponds to the value of the input features. Conversely, amplitude encoding allows encoding $2^n$ input features into the amplitude of a quantum state with $n$ qubits. This method leverages the exponential growth of the Hilbert space with the number of qubits, which is beneficial for handling large-scale data. However, it is often considered less practical for current quantum hardware due to the challenge of efficiently preparing the required quantum state \cite{weigold2020data}.


After data encoding, an ansatz $U_W(\boldsymbol{\theta})$ of $\boldsymbol{\theta}$-parametrized unitaries is applied to the state $\ket{\psi}$. Such unitaries are made of tunable single-qubit rotation gates, which are randomly initialized, and fixed entangling gates. Rotations are used to adequately control the quantum state space, while entanglement allows to create deeply correlated quantum vectors among all the qubits. During the computation, parameters optimization is performed by means of a classical co-processor in an iterative framework, as illustrated in Fig.~\ref{fig:VQC_framework}.

Rotations and entanglement layers may be applied repeatedly to realize more expressive models \cite{abbas2021power}. The list of proposed ansatzes in literature is large and widely studied \cite{cerezoansatz,abbas2021power}. Nevertheless, there is no universal consensus on the optimal design of an ansatz; the choice often depends on the specific problem and the quantum hardware's capabilities. Various types of ansatzes have been explored, including hardware-efficient ansatzes, problem-specific ansatzes, randomized ansatzes and matchgate (theoretical-inspired?).



A measurement operation is performed at the end of the circuit to extract the outcome of the quantum circuit. Typically, qubits are measured in a desired basis to obtain the following prediction:
\begin{equation}	
 f(\mathbf{x}, \boldsymbol{\theta}) = \bra{\phi(\mathbf{x})}U_W(\boldsymbol{\theta})^{\dagger}\hat{O}U_W(\boldsymbol{\theta})\ket{\phi(\mathbf{x})}\,, 
\end{equation}
which corresponds to estimating the expectation value $\langle \hat{O} \rangle$ of an observable $\hat{O}$. Due to the probabilistic nature of quantum mechanics, several shots of the circuits are necessary to get the expected value of the designated observable. The expected value $\langle \hat{O} \rangle$ can be intended as the weighted sum of the eigenvalues, where the weights represent the probabilities that the measured state vector is in the associated eigenstate. 

The outcome $f(\mathbf{x}, \boldsymbol{\theta}) $ of the VQC is then evaluated into a cost function $C(f(\mathbf{x}, \boldsymbol{\theta}))$ and the parameters $\boldsymbol{\theta}$ are properly optimized using a classical processor. Such quantum and classical steps are run iteratively in a cycle to find better $\boldsymbol{\theta}$ parameters at every step. To update $\boldsymbol{\theta}$ and train the VQC, both gradient-free and gradient-based techniques can be used \cite{cerezo2021variational}. Gradients in a PQC are calculated via the parameter-shift rule:
\begin{equation}
     \nabla_{\theta}f(x,\theta) = \frac{1}{2} \left[ f(x,\theta + \frac{\pi}{2}) - f(x,\theta - \frac{\pi}{2}) \right]\,,
\end{equation}
where $f(\mathbf{x}, \theta)$ is the output of the quantum circuit and $\theta$ is the parameter to be optimized.

\subsection{Ising Model}
The Ising model is a fundamental model in statistical mechanics, particularly in the study of phase transitions. In quantum computing, the Ising model serves as a basis for various quantum algorithms and provides insights into quantum phase transitions. The Ising model describes the interactions between spins (or qubits) in a system, influenced by both internal interactions and external magnetic fields. It can be represented by a Hamiltonian, which for the classical Ising model is defined as:
\begin{equation}
\label{eq:ising}
    \mathcal{H}_I = -J \sum_{(j,k)} \sigma_j \sigma_k - \mu \sum_j h_j \sigma_j,
\end{equation}
where $\sigma_j$ represents the spin at site $j$, $J$ indicates the strength of interaction between adjacent spins, $\mu$ is the magnetic moment, and $h_j$ is the external magnetic field acting on spin $j$. The Ising model Hamiltonian is not only significant in classical statistical mechanics but also finds applications in solving graph-based combinatorial optimization problems \cite{barahona1988application}.

The quantum counterpart of the Ising model is the Transverse-field Ising Model (TIM), which plays a crucial role in understanding quantum phase transitions. The TIM Hamiltonian $H_{TIM}$ is defined as follows:
\begin{equation}
\label{eq:ising_TIM}
    \mathcal{H}_\mathrm{TIM} = J \sum_{(j,k)} \sigma_j^z \sigma_k^z - \mu \sum_j h_j \sigma_j^z - \mu \sum_j g_j \sigma_j^x\,,
\end{equation}
where $\sigma_j^z$, $\sigma_j^x$ are Pauli matrices, and $g_j$ represents the longitudinal magnetic field. Through time-evolution operators constructed from the TIM Hamiltonian, it is possible to simulate quantum systems and perform computations relevant to various quantum algorithms.

The importance of the Ising model extends to the realm of QGNNs, as it provides a natural framework for representing and processing graph-structured data. This stems from the ability of the Ising model to encode interactions between nodes (or spins) in a graph, which is analogous to the relationships between data points in graph-based ML tasks. By leveraging the Ising model, QGNNs can effectively utilize quantum correlations and entanglements to enhance the learning and inference processes on graphs. 

Specifically, the Hamiltonian of the Ising model can be employed to represent the adjacency matrix of a graph, where the interaction terms correspond to the edges between nodes. This representation allows QGNNs to incorporate the complex dependencies and connectivity patterns inherent in graph data. Moreover, the optimization problems that arise in training QGNNs, such as finding the optimal set of parameters to minimize a loss function, can be mapped onto Ising-type Hamiltonians. Quantum algorithms designed to solve Ising model instances, such as Quantum Approximate Optimization Algorithm (QAOA) and Variational Quantum Eigensolver (VQE), can thus be adapted for QGNN training.

\subsection{Barren Plateaus}
Training VQCs faces a significant challenge related to the cost function landscape. Specifically, for certain ansatz families, the cost function landscape can be remarkably flat, resulting in gradients that are exponentially small relative to the number of qubits. This phenomenon causes the optimization process to stall and is known as the problem of barren plateaus \cite{mccleanBarrenPlateausQuantum2018}. A cost function \( C(\boldsymbol{\theta}) \) exhibits a barren plateau if, for all trainable parameters \( \theta_i \in \boldsymbol{\theta} \), the variance of the partial derivative of the cost function decreases exponentially with the number $n$ of qubits:
\begin{equation}
\label{eq:barren_plateaus}
    \text{Var}_{\boldsymbol{\theta}}[\partial_i C(\boldsymbol{\theta})] \leq F(n)\,,
\end{equation}
where $F(n) \in O(b^{-n})$ for some constant $b > 1$. The formula in \eqref{eq:barren_plateaus} implies that, on average, the gradient of the cost function will be exponentially small. By applying Chebyshev's inequality, we can see that the probability of the partial derivative $ \partial_i C(\boldsymbol{\theta})$ deviating from its mean (which is zero) by more than a constant $c$, $c > 0$, is bounded by the variance of the gradient $\text{Var}_{\boldsymbol{\theta}}[\partial_i C(\boldsymbol{\theta})]$, as shown in \eqref{eq:barren_plateaus}:
\begin{equation}
\label{eq:barren_plateaus2}
    \text{Pr}[\abs{\partial_i C(\boldsymbol{\theta})} \geq c] \leq \frac{1}{c^2} \text{Var}_{\boldsymbol{\theta}}[\partial_i C(\boldsymbol{\theta})]\,.
\end{equation}

The challenge in training quantum circuits lies in the gradients needed for optimization becoming negligible as the number of qubits increases. Consequently, developing effective strategies to overcome these barren plateaus is crucial for the success of VQAs.

\section{Quantum Graph Neural Networks}\label{sec:qgnns}
 %

\begin{figure*}[!ht]
    \centering
    \includegraphics[width=1\textwidth]{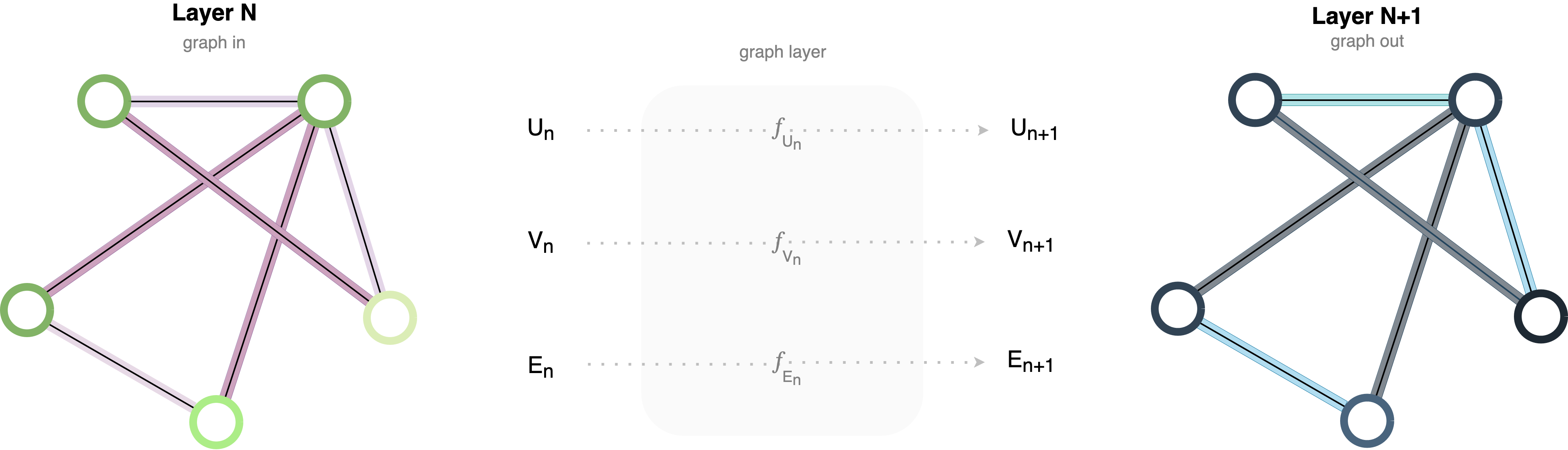}
    \caption{A single layer of a simple GNN. A graph is the input, and each component (V,E,U) gets updated by a MLP (classical computing) or by a VQA (quantum computing) to produce a new graph. Each function subscript indicates a separate function for a different graph attribute at the $n$-th layer of a GNN model.}
    \label{fig:enter-label}
\end{figure*}

The definition a Quantum Graph Neural Network (QGNN) that utilizes a quantum circuit's ansatz to learn and make predictions about the underlying graph is considered a groundbreaking approach in the field of QML. The underlying idea is to leverage the quantum dynamics to embed data and thus introduce a richer feature map, with characteristics that are hard to access for classical methods \cite{innan2024financial}. 

Specifically, in a QGNN the graph topology is encoded in the dynamics through the Hamiltonian of the system and this result implies that it is possible to think about quantum circuits with graph-theoretic properties. Thus, there is a clean way of taking quadratic Hamiltonians and turning them into unitaries (quantum circuits) that preserve the same correspondence to a graph. In general, this kind of unitary describing the whole circuit is very difficult to implement on a quantum computer. However, the Trotter-Suzuki decomposition can be used to approximate it, as shown by Jones et al. \cite{jones2019optimising}. In this sense, QGNNs are intended as VQCs able to learn the underlying dynamics of the system with a variational approach. 

QGNNs were first introduced in 2019 by \cite{verdon2019quantum}. To begin with, the authors introduced the concept of Networked Quantum Systems. They consider a graph $G = (V, E)$, where $V$ is the set of nodes and $E$ the set of edges. In this perspective, one can assign to each node $v \in V$ in the graph a quantum subsystem with Hilbert space $\mathcal{H}_v$, forming a global Hilbert space $H_V \equiv \bigotimes_{v \in V} H_v$ where each of the nodes or subgraphs corresponds to one or several qubits. 
The same holds for the edges $e \in E$, although the authors don't consider them in their work: $H_E \equiv \bigotimes_{e \in E} H_e$, with $H_E \bigotimes H_V$ being the total Hilbert space for the graph. Therefore, it was defined \textit{quantum network} a setup where the edges of the graph dictate the communication between the nodes sub-spaces \cite{kimble2008quantum,qian2019heisenberg}: couplings between degrees of freedom on two different nodes are allowed if there is an edge connecting them. The topology of this quantum network directly depends on the given graph $G$.

Since QGNNs fall under the category of VQAs, a known issue in training it is the presence of the so-called \textit{barren plateaus}, i.e. regions in the loss landscape where the gradients of the cost function become extremely small or vanish, making it difficult for optimization algorithms to make meaningful updates to the model's parameters \cite{mcclean2018barren}, in a way similar to vanishing gradient for DNNs. QGNNs were devised as a mean of imparting structural information to VQAs in order to ameliorate the presence of such barren plateaus. 

As proposed in \cite{verdon2019quantum}, the most general QGNN ansatz is a PQC on a network that consists of a sequence of $Q$ different Hamiltonian evolutions, with the whole sequence repeated $P$ times:
\begin{equation}
\label{eq:qgnn_ansatz}
    U_\mathrm{QGNN}(\boldsymbol{\eta},\boldsymbol{\theta}) = \prod_{p=1}^P \prod_{q=1}^Q e^{-i\eta_{pq}\hat{H}_q(\boldsymbol{\theta})}\,\,,
\end{equation}
where the product is time-ordered, the $\eta$ and $\theta$ are variational (trainable) parameters, and the Hamiltonians $\hat{H}_q(\boldsymbol{\theta})$ can generally be any parameterized Hamiltonians whose topology of interactions is that of the problem graph:
\begin{equation}
\label{eq:qgnn_ansatz_2}
	\begin{split}
    \hat{H}_q(\boldsymbol{\theta}) \equiv &\sum_{(j,k) \in E} \sum_{r \in \mathcal{I}_{jk}} W_{qrjk}\hat{O}_{j}^{(qr)} \otimes \hat{P}_k^{(qr)}\, + \\
																	        &\,\,\,\,\sum_{v \in V} \sum_{r \in \mathcal{J}_v} B_{qrv}\hat{R}_j^{(qv)}\,.
	\end{split}
\end{equation}

Here the $W_{qrjk}$ and $B_{qrv}$ are real-valued coefficients which can generally be independent trainable parameters, forming a collection $\boldsymbol{\theta} \equiv \cup_{q,j,k,r}\{W_{qrjk}\} \cup_{q,v,r}\{B_{qrv}\}$. The operators $\hat{R}_j^{(qv)}$, $\hat{O}_j^{(qr)}$, and $\hat{P}_j^{(qr)}$ are Hermitian operators which act on the Hilbert space of the $j$-th node of the graph. The sets $\mathcal{I}_{jk}$ and $\mathcal{J}_{v}$ are index sets for the terms corresponding to the edges and nodes, respectively. To make compilation easier, one can enforce that the terms of a given Hamiltonian $\hat{H}_q$ commute with one another, but different $\hat{H}_q$ need not commute.

In order to make the ansatz more amenable to training and avoid the barren plateaus/no free lunch problem, there is the need to add some constraints and specificity. To that end, this paper propose more specialized architectures where parameters are tied spatially (convolutional) or tied over the sequential iterations of the exponential mapping (recurrent).

Based on the taxonomy introduced by Wu et al. \cite{wu2020comprehensive} (classical counterpart), in this paper QGNNs are categorized into: Quantum Recurrent Graph Neural Networks (QRecGNNs), Quantum Convolutional GNNs (QConvGNNs), 
and Quantum Spatial–Temporal GNNs (QSTGNNs). Moreover, leveraging the concept of equivariance taken from geometric ML, they are also introduced Equivariant Quantum Graph Circuits (EQGCs).

\subsection{Quantum Recurrent Graph Neural Networks}
QRecGNNs are the quantum analogue of classical RecGNNs, and a subclass of the more general QGNN ansatz. Authors in \cite{verdon2019quantum} define QRecGNNs as ansatzes described by \eqref{eq:qgnn_ansatz}, where the temporal parameters are the same for each iteration, so that $\eta_{pq} \rightarrow \eta_q$. This is similar to classical RecGNNs, where parameters are shared over sequential applications of the recurrent map. As $\eta_q$ acts as a time parameter for Hamiltonian evolution under $\hat{H}_q$, the QGRNN ansatz can be viewed as a Trotter-based quantum simulation of an evolution $e^{-i\Delta\hat{H}_{eff}}$ under the Hamiltonian $H_{eff} = \Delta^{-1} \sum_q \eta_q \hat{H}_q$ for a time step of size $\Delta = |\boldsymbol{\eta}|_1 = \sum_q |\eta_q|$.

This approach is tailored to learn effective dynamics occurring on a graph. In other terms, QGRNN is a PQC which maps the underlying graph structure into a quantum circuit, consisting of $Q$ different Hamiltonian evolution sequences with the entire sequence repeated $P$ times, and is generally defined as follows:
\begin{equation}
    U_\mathrm{QRecGNN}(\boldsymbol{\eta},\boldsymbol{\theta}) = \prod_{p=1}^P \prod_{q=1}^Q e^{-i\eta_{q}\hat{H}_q(\boldsymbol{\theta})}\,.
    \label{eq:qgrnn_ansatz}
\end{equation}

Moreover, for the QRecGNN to effectively handle graphs, the structure of the quantum circuit is vital. It is important to define a quantum circuit architecture that encompasses all essential graph operations and transformations. To model graph operations such as node updating and message passing, it is necessary to determine the quantity and type of quantum gates or operations used \cite{10391361}.

\subsection{Quantum Convolutional Graph Neural Networks}
Classical Convolutional Graph Neural Networks (ConvGNNs) fundamentally rely on permutation invariance, meaning the model should remain unchanged under any permutation of the nodes. This is analogous to how traditional convolutional transformations are invariant to translations. In quantum computing, permutation invariance imposes a restriction on the Hamiltonian, necessitating the use of global trainable parameters instead of local ones. As seen in \eqref{eq:qgnn_ansatz} for Quantum Graph Convolutional Neural Networks (QConvGNNs), the parameters $\boldsymbol{\theta}$ become tied over indexes of the graph, with $W_{qrjk} \rightarrow W_{qr}$ and $B_{qrv} \rightarrow B_{qr}$. This structure is reminiscent of the Quantum Alternating Operator Ansatz, a generalized form of the Quantum Approximate Optimization Algorithm, where variational parameters are shared across the circuit.

In \cite{zheng2021quantum}, a QConvGNN model is proposed featuring state preparation, quantum graph convolution, quantum pooling, and quantum measurements. Here, node features are encoded into a quantum register per node, and connectivity is represented by $\ket{0}$ or $\ket{1}$ on the node-pair-representing qubits. Convolution is implemented as controlled two-node unitaries, and pooling as measurement-conditioned unitary operations. This model mimics classical GCNNs using Parameterized Quantum Circuits (PQCs) and it is designed to solve graph-level classification tasks. The design procedure includes:
(i) Encoding graph data into quantum states via amplitude encoding, as in \eqref{eq:amplitude_encoding}; (ii) using parameter-based universal quantum gates to construct the QConvGNN post-amplitude encoding; (iii) outputting classification results through the quantum measurement block; (iv) training the QConvGNN model in a Variational Quantum Algorithm (VQA) fashion using a specific dataset.

The quantum graph convolutional layer aggregates neighbor nodes to the current node, with the number of layers representing the order of node aggregation. Parameter sharing within the same layer is a key characteristic. To implement an arbitrary unitary two-qubit operation with minimal gates, the design leverages \cite{shende2004minimal}. The quantum pooling layer, distinct from the convolutional layer, reduces feature dimensions via intermediate quantum measurements, analogous to classical pooling. It includes a CNOT gate, parameterized $R_y$ and $R_z$ gates, and their inverses.
The proposed model effectively captures node connectivity and learns hidden layer representations of node features. 

In another work \cite{10499715}, Zhang et al. propose a QConvGNN model for non-Euclidean data, implemented on PQCs. This model includes quantum encoding, a quantum graph convolutional layer, a quantum graph pooling layer, and network optimization. Quantum amplitude encoding is used for converting classical feature vectors into quantum states. The graph convolutional layer embeds non-Euclidean data's topological structure into a hierarchical network, while the pooling layer employs a hierarchical mechanism to reduce data size and improve training efficiency. The Quantum Stochastic Gradient Descent (QSGD) algorithm optimizes parameters, leveraging the parameter shift rule and analyzing the barren plateau phenomenon to ensure effective training.

The hybrid quantum-classical architecture combines quantum encoding, the QNN model, and a classical optimizer. The optimizer updates parameters via gradient descent, feeding them back into the quantum circuit.
In the paper \cite{hu2022design}, a novel method for implementing Graph Convolutional Neural Networks (GCNs) using quantum circuits, referred to as QuGCN, is proposed. This approach leverages the inherent parallelism of quantum computing to address the computational and memory limitations faced by classical GCNs as the graph size and number of features increase.

The primary challenge addressed by QuGCN is integrating graph topology information into quantum circuits while maintaining the learnable parameters characteristic of classical GCNs. Givens rotations encode graph information, and VQCs handle learnable parameters. 
In classical GCNs, the adjacency matrix represents the graph structure and facilitates the propagation of node information through the network. QuGCN employs Givens rotations to mimic this function in a quantum circuit. The fundamental representation of a Givens rotation is given by:
\begin{equation}
G_{N}(i, j, \theta) = 
\begin{pmatrix}
1 & \cdots & 0 & \cdots & 0 & \cdots & 0 \\
\vdots & \ddots & \vdots & & \vdots & & \vdots \\
0 & \cdots & \cos \theta & \cdots & \sin \theta & \cdots & 0 \\
\vdots & & \vdots & \ddots & \vdots & & \vdots \\
0 & \cdots & -\sin \theta & \cdots & \cos \theta & \cdots & 0 \\
\vdots & & \vdots & & \vdots & \ddots & \vdots \\
0 & \cdots & 0 & \cdots & 0 & \cdots & 1 
\end{pmatrix}
\end{equation}
where $\cos\theta$ and $\sin\theta$ are placed at the $i$-th and $j$-th rows and columns, respectively.

Node features are encoded using amplitude encoding as in \eqref{eq:amplitude_encoding}, significantly reducing quantum resource requirements compared to methods requiring $O(N)$ or more qubits. The QuGCN framework integrates VQCs for learnable parameters, with node feature propagation analogous to classical GCNs' layer-wise propagation rule:
\begin{equation}
    H^{(l+1)} = \sigma\left( \tilde{D}^{-\frac{1}{2}} \tilde{A} \tilde{D}^{-\frac{1}{2}} H^{(l)} W^{(l)} \right)\,.
\end{equation}
Here, $\tilde{A}$ is the adjacency matrix with added self-loops, $\tilde{D}$ is the degree matrix, $H^{(l)}$ represents node features at layer $l$, and $W^l$ is the learnable weight matrix. In QuGCN, these matrices are encoded using Givens rotations within the quantum circuit.

QuGCN presents a significant step towards leveraging quantum computing for graph-structured data, addressing computational complexity and memory limitations of classical approaches. The use of Givens rotations and amplitude encoding within a VQC framework provides a scalable and efficient solution, poised to take advantage of near-term quantum devices in the NISQ era.

Moreover, in \cite{bai2021learning}, Bai et al. introduce a Quantum Spatial Graph Convolutional Neural Network (QSGCNN) for graph classification tasks. This model integrates quantum information theory, specifically continuous-time quantum walks, to enhance graph structure representation and learning. Quantum walks provide richer information about node connections and transitions due to their quantum properties, with the average mixing matrix capturing transition probabilities in a quantum system.

The methodology begins with aligning input graphs to a prototype graph structure, ensuring no vertex information is discarded. The resulting grid structure is processed by standard convolutional neural networks (CNNs). The QSGCNN architecture includes multiple quantum spatial graph convolution layers followed by traditional CNN layers, leveraging both spatial graph information and global graph topology for an end-to-end deep learning architecture.

Finally, the quantum aspect of the methodology proposed in \cite{shah2021quantum}, focuses on the integration of the Quantum Fisher Information Matrix (QFIM) into the QconvGNN. In fact, QFIM is central to estimating quantum parameters, particularly using the Cramér-Rao bound in quantum mechanics. QFIM interacts with key quantum concepts such as the quantum geometric tensor and entanglement witnesses, making it crucial for high-sensitivity measurements and understanding physical structures.
Specifically, the QConvGNN model incorporates QFIM to enhance the extraction of multi-scale vertex features in graph data. QFIM helps in propagating vertex information effectively across the graph, leveraging quantum theories to improve the model's performance in text classification tasks. 
It was observed that, by using QFIM, the QGCN can achieve more precise and robust feature extraction compared to classical methods.

\subsection{Quantum Spectral Graph Convolutional Neural Networks}
In \cite{verdon2019quantum}, the authors draw inspiration from the continuous-variable quantum approximate optimization ansatz introduced in \cite{verdon2019quantum2} to create the Quantum Spectral Graph Convolutional Neural Network (QSGCNN). This model reinterprets Laplacian-based ConvGNNs proposed by \cite{kipf2016semi} in the Heisenberg picture, using continuous-variable quantum computers designed for processing continuous variables like position and momentum. These computers operate in an infinite-dimensional Hilbert space, capable of performing operations related to position or momentum in a quantum system. However, digital quantum computers, which operate with qubits in a finite set of discrete states, can only emulate this behavior by using multiple qubits, leading to inefficiency or infeasibility for modern quantum devices. Therefore, this paper does not consider this type of QGNN, focusing instead on digital quantum computers.

\subsection{Quantum Spatial–Temporal GNNs}
Quantum Spatial Temporal GNNs represent a cutting-edge convergence of quantum computing and GNNs, specifically tailored to handle the intricacies of spatial-temporal data. The methodology behind Quantum Spatial–Temporal GNNs, proposed by Qu et al. \cite{qu2022temporal}, begins with the transformation of classical data into quantum states using quantum circuits. This transformation is followed by a temporal analysis based on the Schrödinger approach, where data dynamics are modeled as quantum walks, capturing the evolution of quantum states over time. The spatial dimension of data is then processed using a QConvGNN, which integrates temporal features with the graph structure to make predictions. This method adeptly leverages the time evolution properties of quantum wave functions and parametric quantum circuits to capture the temporal and spatial characteristics of data.

To demonstrate the robustness of the model, we examine the impact of perturbations caused by four types of quantum noise: bit flip noise, phase flip noise, amplitude damping noise, and depolarizing noise. The Kraus operators for these noises are presented in \eqref{eq:bit_flip}, \eqref{eq:phase_flip}, \eqref{eq:amplitude_damping} and \eqref{depolarizing_noise}, respectively:

\begin{equation}
    E_{0} = \sqrt{\gamma_\mathrm{BF}} I\,, \quad E_{1} = \sqrt{1 - \gamma_\mathrm{BF}} \sigma_{x}\,.
    \label{eq:bit_flip}
\end{equation}

\begin{equation}
    E_{0} = \sqrt{\gamma_\mathrm{PF}} I\,, \quad E_{1} = \sqrt{1 - \gamma_\mathrm{PF}} \sigma_{z}\,.
    \label{eq:phase_flip}
\end{equation}

\begin{equation}
    E_{0} = \begin{pmatrix}
    1 & 0 \\
    0 & \sqrt{\gamma_\mathrm{AD}}
    \end{pmatrix}\,, \quad E_{1} = \begin{pmatrix}
    0 & \sqrt{1 - \gamma_\mathrm{AD}} \\
    0 & 0
    \end{pmatrix}\,.
    \label{eq:amplitude_damping}
\end{equation}

\begin{equation}
    \begin{aligned}
        &E_{0} = \sqrt{\gamma_\mathrm{D}} I\,, \quad &E_{1} = \sqrt{\frac{1 - \gamma_\mathrm{D}}{3}} \sigma_{x}\,,\\
        &E_{2} = \sqrt{\frac{1 - \gamma_\mathrm{D}}{3}} \sigma_{z}\,, \quad &E_{3} = \sqrt{\frac{1 - \gamma_\mathrm{D}}{3}} \sigma_{y}\,.
    \end{aligned}
    \label{depolarizing_noise}
\end{equation}
In these equations, $\gamma_\mathrm{BF}$, $\gamma_\mathrm{PF}$, $\gamma_\mathrm{AD}$, and $\gamma_\mathrm{D}$ represent the noise parameters, each ranging from 0 to 1, corresponding to bit flip noise, phase flip noise, amplitude damping noise, and depolarizing noise, respectively. $I$ denotes the identity matrix, while $\sigma_{x}$, $\sigma_{y}$, and $\sigma_{z}$ are the Pauli matrices.

The promise of Quantum Spatial–Temporal GNNs lies in its theoretical advantages over classical models. Quantum algorithms can process large datasets more efficiently, potentially reducing computational overheads in handling complex spatial-temporal problems. Moreover, the quantum approach allows for a more precise extraction of both spatial and temporal features, potentially leading to superior predictive performance. The scalability of quantum models also holds promise for application to larger and more complex tasks.

However, several challenges remain. Current quantum hardware is still in its developmental stages, with limited qubits and high error rates that could affect the practical implementation of Quantum Spatial–Temporal GNNs. While the theoretical scalability is promising, actual implementation on larger model, such as traffic networks, requires more advanced quantum processors and efficient quantum error correction methods. Additionally, the complexity of developing and training Quantum Spatial–Temporal GNNs models poses significant barriers, requiring specialized expertise in both quantum computing and GNNs.

\subsection{Quantum GNNs from Quantum Graph States}
A first defined graph-structured quantum data was proposed in \cite{beer2023quantum}. Given a graph, each node corresponds to a quantum state, and there is a link between two quantum states if they are within a certain information theoretical distance of each other. Then, one can define a supervised learning scenario where each node should be mapped to a certain labeled quantum state. This work provides loss function designs and training methods for this task using Dissipative Quantum Neural Networks (DQNNs). Further developments in this regard have been recently proposed in \cite{daskin2024}, showing that QGNNs can be used as parameterized quantum circuits to represent neural networks or to build GNNs on quantum computers.

\subsection{Equivariant Quantum Graph Circuits}
\begin{figure}[!ht]
    \centering
    \includegraphics[width=0.9\columnwidth]{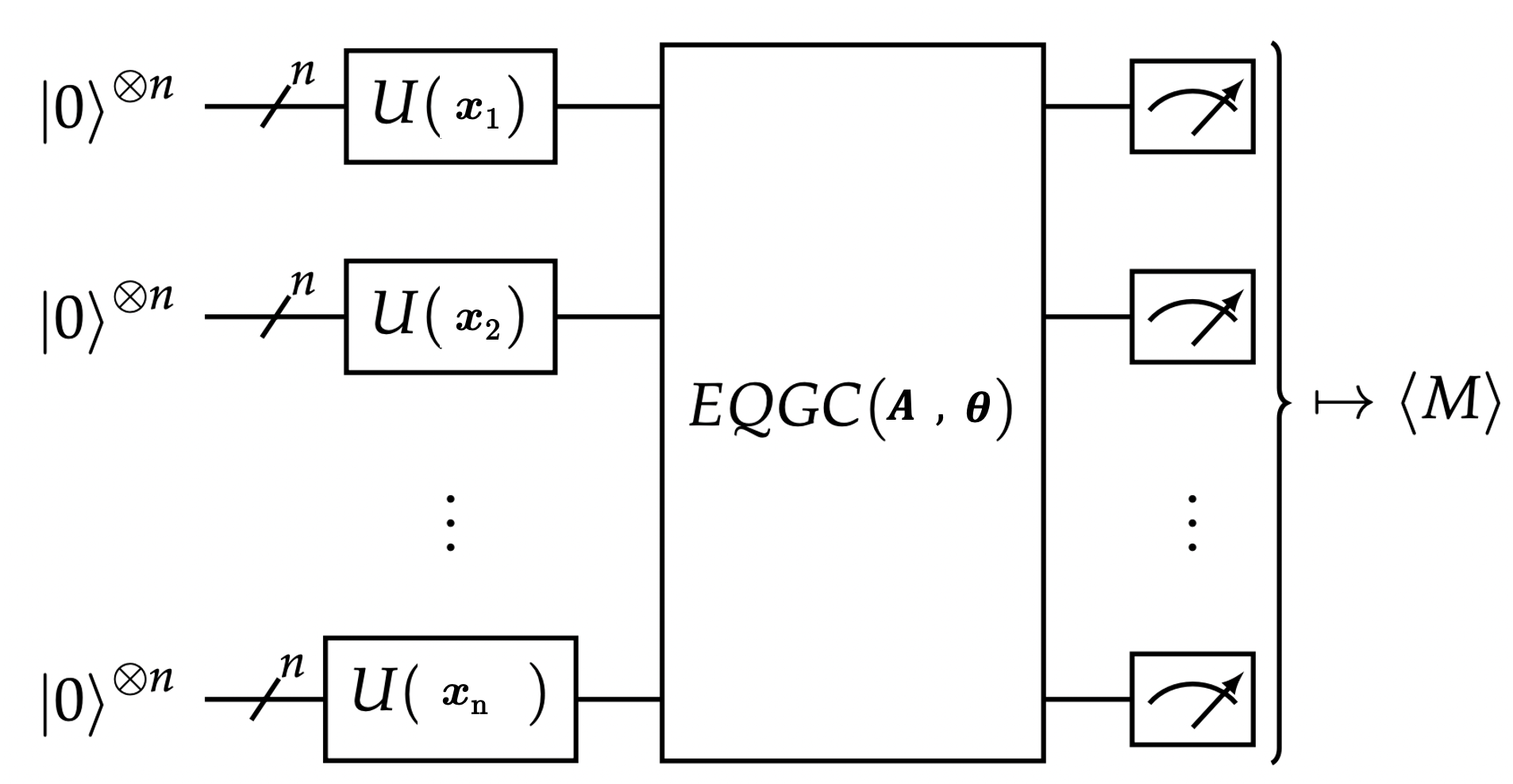}
    \caption{The EQGC framework. Given an input graph with $n$ nodes, Adjacency matrix $\mathbf{A}$, and node features $\mathbf{x}_i, i = 1\dots n$, the circuit drawn above is used in the computation. The unitary EQGC is equivariant on the permutation of nodes and is trainable on parameters $\boldsymbol{\theta}$, while the measurement is invariant on the permutation of nodes \cite{mernyei2023equivariant}.}
    \label{fig:eqgnn}
\end{figure}

An implementation of Equivariant Quantum Graph Circuits (EQGCs) was developed by Mernyei et al. \cite{mernyei2023equivariant}, drawing on the concept of equivariance from geometric ML. Specifically, given input graph data made up of a tuple of nodes, links, and node features, some fixed number of qubits are assigned to each node. Next, the node features are encoded into quantum states by applying a parameterized unitary onto the corresponding qubits. Then, a node-permutation-equivariant quantum circuit is applied, i.e. a unitary matrix which is parametrized and depends in some way on the Adjacency matrix of the graph. Lastly, a node-permutation-invariant measurement is applied and then the outputs are post-processed, for example by applying a parametrized classical function to the outputs of the circuit. In this case, equivariance to permutation means that the permutation can commute with the quantum circuit. The node-permutation-invariant measurement is simple to design, such as the average of the expectation values of a node-local observable over all nodes. This framework can be visualized as the circuit in Fig.~\ref{fig:eqgnn}.

There are two main ways of constructing an EQGC as suggested in \cite{mernyei2023equivariant}. One is the Equivariant Hamiltonian Quantum Graph Circuit (EHQGC), where the QNN is made up of unitaries with rotation-generating Hamiltonian operators that have the same topology as the input graph, which is an approach very similar to the one proposed by \cite{verdon2019quantum}. Another method is the one introduced in \cite{ryu2023quantum}, called Equivariantly Diagonalizable Unitary Quantum Graph Circuit (EDU-QGC). EDU-QGCs are made up of node layers and link layers. Each node layer is made up of a node-local unitary operator that acts on all the nodes. Each link layer is made up of Equivariantly Diagonalizable Unitaries (EDUs) acting between two nodes connected by a link. An EDU is defined as a unitary acting on two nodes that can be decomposed as:
\begin{equation}
    \mathrm{EDU} = (\mathbf{V}^{\dagger} \otimes \mathbf{V}^{\dagger})\mathbf{D}(\mathbf{V} \otimes \mathbf{V})\,,
\end{equation}
where $\mathbf{V}$ is a unitary operator which acts on one node, while $\mathbf{D}$ is a diagonal unitary that acts on two nodes. 

Since the EDU commutes with the SWAP operator and a copy of itself acting on other qubits, the link layer is equivariant on the permutation of the nodes. Overall, EDU-QGCs can still be represented as EHQGCs. Finally, a recent research \cite{mernyei2023equivariant} provides also some expressibility results about EDU-QGCs, revealing that EDU-QGCs can approximate any real-valued function over bounded graphs and can pass the 1-Weisfeiler-Lehman Isomorphism test, which deterministic classical MPNNs cannot pass. 

One limitation of the previous approaches is that they can only handle small graphs in this moment, due to the difficulty in using large quantum circuits with many qubits. In \cite{ai2022decompositional}, it was proposed a more comprehensive quantum graph model to not only achieve the goal of simulating GNNs in a quantum framework, but also proposed a strategy to solve the lack of available qubits. The key in their method was to use subgraph decomposition and CNOT gates to handle and decompose the topology of a given graph. For a given graph with $n$ nodes, they split the whole graph into $n$ subgraphs. Thus, each subgraph consisted of a node with its neighbors, represented by qubits equal to the number of subgraph's nodes and followed by a series of trainable parameterized quantum gates. This way, they could entangle information by applying CNOT gates to each pair of nodes in a subgraph. Finally, they combined all subgraphs and got the graph representation for the final prediction task.

Furthermore, the article referenced in \cite{forestano2024comparison} provides an in-depth analysis of the approaches used for Equivariant Quantum Graph Neural Networks (EQGNN) in the context of Large Hadron Collider (LHC) data analysis, comparing them to both classical and quantum neural networks. As previously highlighted, EQGNNs maintain permutation equivariance by symmetrically aggregating elements of the quantum state before making predictions.

Both QGNNs and EQGNNs rely on a layered structure with parameterized Hamiltonians and unitary operators to evolve quantum states. EQGNNs add an aggregation layer to preserve equivariance.
In evaluating a binary classification task (identifying jet particles originating from quarks or gluons), EQGNNs demonstrated superior performance compared to their classical and non-equivariant quantum counterparts, achieving a 7.29\% improvement in performance.
Future developments include exploring various network forms, such as the introduction of attention components and modifying the structure of quantum graph layers, which could further enhance the performance of both classical and quantum neural networks.

In summary, this paper highlights how EQGNNs leverage fundamental symmetries in the data and the advanced computational capabilities of quantum systems to improve high-energy data analysis, although practical implementations still face significant challenges, such as implementation on quantistic hardware and the number of available qubits.

\section{Applications and Potential \\for Quantum Advantage}\label{sec:applications}

QGNNs have demonstrated numerous advantages over traditional GNNs, showcasing their versatility across various and diverse applications. Here, we highlight some of the most significant applications where QGNNs have fully exhibited their potential, illustrating their effectiveness in addressing complex challenges and improving performance in a range of fields.

\subsection{High-Energy Physics}
A fundamental application of QGNNs has been demonstrated in particle track reconstruction \cite{Amrouche_2019}. At the European Organisation for Nuclear Research (CERN), particles accelerated in the Large Hadron Collider (LHC) collide in bunches, producing new particles that scatter in all directions and generate signals as they pass through tracking detectors. Traditional tracking algorithms, used to distinguish these signals and identify particle trajectories, face scalability challenges, making rapid track reconstruction difficult, especially with the upcoming High Luminosity LHC upgrade, which will increase the particle density in the beam \cite{https://doi.org/10.5170/cern-2015-005.1}. 

To address this, \cite{zenodo.4088474} and \cite{tuysuz2021hybrid} propose leveraging QGNNs to enhance the speed and efficiency of particle track reconstruction. Their hybrid model incorporates an Input Network to expand input data dimensions, followed by Edge and Node Networks that combine classical and quantum layers. This flexible architecture allows for variations in hidden dimension size \( N_D \), the number of qubits, and the quantum circuit used. The results, analyzed with varying hyperparameters, show that this hybrid model performs comparably to classical models and therefore appear promising, especially considering the simplicity of the model. This suggests the possibility of achieving a more pronounced quantum advantage in the use of QGNNs in particle track reconstruction, perhaps by investigating more sophisticated encoding.

Another promising application of QGNNs in High Energy Physics is in identifying the type of elementary particles that initiate jets—collimated sprays of particles formed through hadronization. In \cite{chen2024jetdiscriminationquantumcomplete}, a QCGNN is proposed for this task. Using datasets as MADGRAPH5 \cite{Alwall_2014}, PYTHIA8 \cite{bierlich2022comprehensiveguidephysicsusage}, and DELPHES \cite{ovyn2010delphesframeworkfastsimulation}, which simulate hard scattering processes, QCD hadronization, and detector responses, the jets are clustered with the $k_t$ algorithm  with $R=0.8$ and FASTJET \cite{de_Favereau_2014}. Each jet is represented as a complete graph with nodes corresponding to particles and energy flow information as node features. Compared to classical message-passing GNNs (MPGNN), the QCGNN shows significant advantages. While the predictive power of the MPGNN and QCGNN are comparable with similar parameter counts, the QCGNN exhibits more stable training, faster convergence, and reduced fluctuations.

\subsection{Molecular Chemistry and Biology} 
The use of GNNs for predicting material properties has become increasingly essential due to the high computational costs associated with traditional laboratory techniques, which grow with the size of the molecules considered \cite{https://doi.org/10.1002/wcms.1290}. With the growing availability of datasets \cite{https://doi.org/10.1063/1.4812323}, data-driven approaches are in higher demand. However, GNNs encounter limitations such as insufficient training data for exotic compounds and convergence issues with message-passing schemes \cite{liu2021eignn}. QGNNs present a promising alternative, leveraging QML to improve performance in scientific and molecular engineering applications \cite{2023_Bhatia, Mensa_2023}.

In predicting molecular properties, QGNNs can significantly enhance performance by mapping molecular structures into higher-dimensional spaces and capturing long-range correlations through quantum contextuality. For instance, \cite{Vitz2024HybridQG} explores a Hybrid Quantum-Classical Convoluted Graph Neural Network (HyQCGNN) designed to predict the formation energy of perovskites using a specific dataset \cite{C2EE22341D}. This hybrid model employs the GENConv model \cite{li2020deepergcnneedtraindeeper} for classical processing of node and edge features, which are then input into a quantum circuit. Although the \(R^2\) value from the HyQCGNN is lower compared to a custom classical model and XGBoost \cite{10.1145/2939672.2939785}, the results are promising, indicating that further improvements in quantum graph embeddings could lead to more pronounced quantum advantages.

In another study \cite{37374486}, QGNNs are utilized to infer the the highest occupied molecular orbital-lowest occupied molecular orbital (HOMO-LUMO) energy gap using the QM9 dataset \cite{23088335}.The model chosen is the EDU-QGC, based on what was previously described in \cite{mernyei2023equivariant}. The design of the model is divided into three parts: quantum encoding, link feature processing, and readout functions. The encoding uses RY and RZ gates with three different encoding methods based on features like Atomic Number, Number of Hydrogens, Aromaticity, and Hybridization. The model, compared to the version presented in \cite{mernyei2023equivariant},  is enhanced to process link features, with two variations of the EDU: a default with a single qubit unitary and a simpler version using RY gates. 

In order to maintain node permutation equivariance, a fixed order is imposed on EDUs. Finally, two readout functions, local and global, are used, which are measurements invariant to the permutation of the qubits. In addition to this model, a hybrid model is also proposed, where the only difference is in the introduction of an artificial neural network to optimize encoding angles.  Finally, an additional variation was proposed by adding a master node to the molecular graph, which is connected to all nodes. The results obtained were analyzed by varying the encoding, the EDU-QGC layers, and the readout functions, and compared with a custom classical model designed to be as similar as possible to the quantum one and with two state-of-the-art models, ALIGNN \cite{Choudhary_2021} and SphereNet \cite{liu2022spherical}.

The results show that the best encoding method combines Atomic Number and Number of Hydrogens, but it still underperforms compared to the hybrid model. This highlights the importance of incorporating an additional neural network to derive encoding angles from a larger set of features. The results show also that quantum models, including baseline, hybrid, and those with a master node, outperformed simpler classical models with similar parameters and demonstrated faster convergence. However, more advanced classical models still outperformed the quantum models, indicating  the need for further development in quantum approaches for materials search problems. 

QGNNs also hold promise in drug discovery by analyzing proteins and their roles in disease mechanisms \cite{article}. Although ML has been critical in this area \cite{Vamathevan2019ApplicationsOM}, QML offers enhanced capabilities for complex, high-dimensional challenges \cite{10.1147/JRD.2018.2888987, 8585034}. A hybrid model that integrates 3D-CNN and SG-CNN with a QNN has shown superior performance on the PDBbind dataset \cite{wwPDB2018} compared to classical models, with improvements in metrics such as RMSE, MAE, \(R^2\), Pearson coefficient and Sperman coefficient. Additionally, again the hybrid model exhibits smoother and more stable convergence.

In cancer biology, QGNNs are applied to analyze cancer cell interactions, which is crucial for selecting appropriate treatments. Traditional GNNs face challenges like over-smoothing and over-squashing \cite{chen2019measuringrelievingoversmoothingproblem, topping2022understandingoversquashingbottlenecksgraphs}, but QGNNs offer potential solutions. For example, \cite{ray2023hybridquantumclassicalgraphneural} enhances HACT-NET \cite{Pati2021HierarchicalGR} with a VQC to process output dimensions. The tests conducted on the Breast Cancer Subtyping (BRACS) dataset \cite{Pati2021HierarchicalGR}, and a comparison with the classical model built similarly but with an MLP instead of the VQC, show that the hybrid model performs comparably or slightly better than classical models for larger output dimensions, highlighting that QGNNs can be a significant tool in practical scenarios.

\begin{figure*}[!ht]
		\centering
    \begin{subfigure}[b]{0.45\textwidth}
    \centering
    \includegraphics[height=60mm, keepaspectratio]{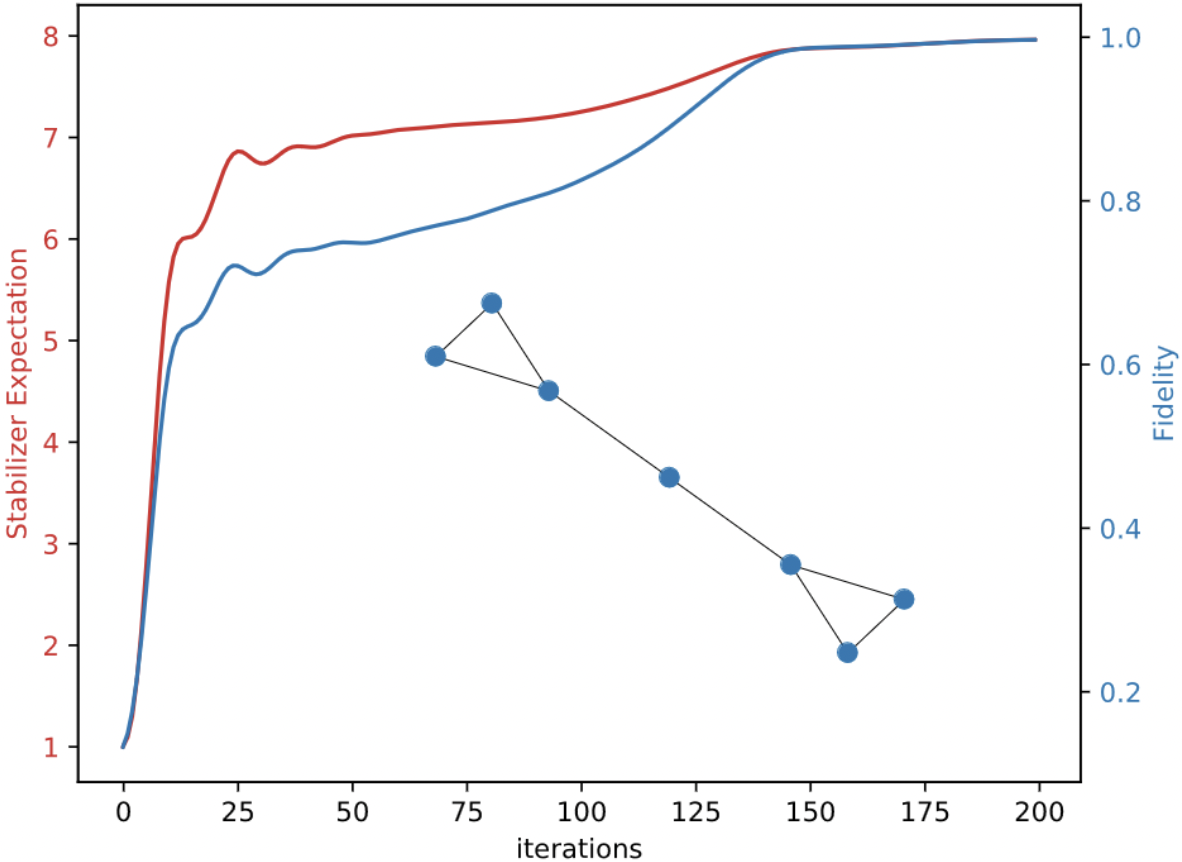}
    \caption{\label{fig:exp2_a}}
    \end{subfigure}
\quad\quad
    \begin{subfigure}[b]{0.45\textwidth}
    \centering
    \includegraphics[height=60.5mm, keepaspectratio]{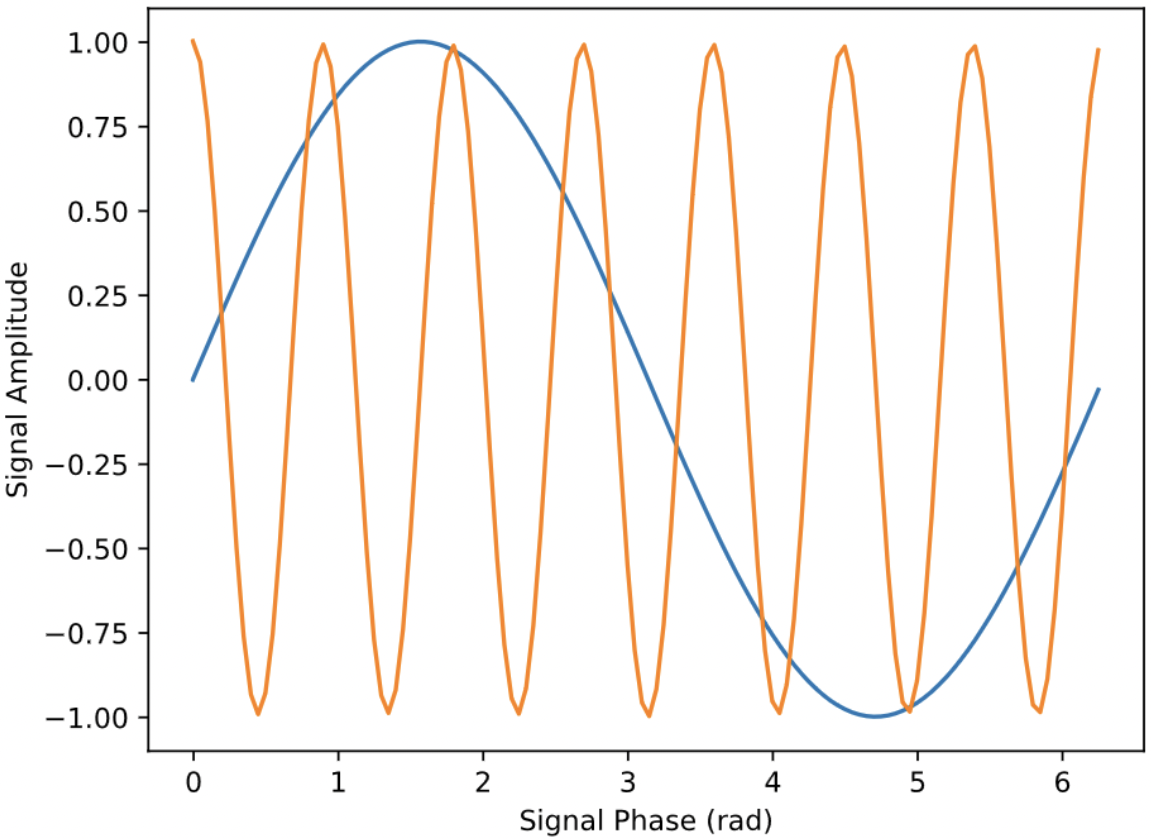}
    \caption{\label{fig:exp2_b}}
    \end{subfigure}
    \caption{(\subref{fig:exp2_a}) The plot shows the progress of stabilizer Hamiltonian expectation and fidelity during training iterations. A visual representation of the quantum network's structure is included in the corner. (\subref{fig:exp2_b}) A quantum phase kickback test is conducted on the acquired GHZ state. Remarkably, for a 7-node network, we can observe a significant $7\times$ increase in the frequency of Rabi oscillations. This demonstrates that they have successfully achieved the Heisenberg limit of sensitivity for their quantum sensor network, i.e. it means that they have achieved a substantial increase in the rate at which quantum transitions in energy levels occur, and this level of sensitivity is at the theoretical maximum allowed by quantum mechanics \cite{verdon2019quantum}.}
    \label{fig:exp2}
\end{figure*}

\subsection{Complex Systems}
The use of QGNNs has shown significant promise in analyzing complex systems, particularly in applications where classical GNNs struggle to capture intricate spatio-temporal dynamics. In earth observation tasks like long-range forecasting of the Ocean Niño Index (ONI) and in traffic management systems, such as traffic congestion prediction (TCP) and traffic collision avoidance systems (TCAS), QGNNs demonstrate a notable advantage by effectively handling complex dependencies and dynamic changes.

For instance, intelligent transportation systems generate vast amounts of traffic data, making it essential to predict traffic conditions accurately \cite{8771378}. The problem of TCP has therefore become popular research for having an intelligent transportation system \cite{JIANG2022108544}. While GNNs have been effective in traffic prediction due to their ability to process non-Euclidean data, they fall short in capturing the full complexity of spatio-temporal characteristics \cite{9515359}. For this reason, efforts have been made to leverage QGNNs also for traffic congestion prediction. The authors of \cite{9882032} propose a new architecture, called the Temporal Spatial Quantum Graph Convolutional Neural Network (TS-QGCNN). Initially, they analyze traffic data in quantum states using a Schrödinger approach to have a closed-form solution of the problem in the time dimension. Then, the QGCNN is applied to capture the spatial features, so the model incorporates not only the spatial topology of traffic data but also its dynamic changes. The results obtained with this model are promising and represent a concrete use of a quantum algorithm to address traffic congestion prediction problems. Once again, classical models achieve better results, but the possibility of significantly improving the process of transforming classical traffic data into quantum data may eventually lead to better outcomes.

In another example \cite{Novotny2021QuantumTR}, the same architecture used in \cite{tuysuz2021hybrid} for particle track reconstruction problems can also have potential applications in the industrial world, such as in the sector of reconstructing airplane routes. Here, the basic idea remains the same, but the particles from \cite{tuysuz2021hybrid} now correspond to airplanes. The objective is thus to optimize airport capacities and improve predictions of potential collisions by the TCAS. In fact, TCAS has become crucial for potential collisions after two aircraft collided, and the use of QGNNs in this context can bring numerous benefits.

Similarly, in earth observation, GNNs have begun to be used as essential tools for capturing global-scale dependencies. It has become crucial to learn the characteristics of spatially distant relationships of earth phenomena that also vary over time. An example of this is the task of seasonal and long-range forecasting of the ONI, where a challenge is the presence of so-called El Niño fluctuations \cite{Sarachik_Cane_2010}, a recurring climatic phenomenon characterized by an abnormal warming of the central and eastern tropical Pacific Ocean. To address this task, the authors of \cite{mauro2024hybrid} propose a hybrid network called HQGCNN, where essentially a GCNN is combined with a Quantum MLP layer. The results obtained demonstrate the necessity of incorporating quantum elements within GNNs. Not only does the performance improve compared to other state-of-the-art models \cite{cachay2021world}, but there is also a significant reduction in training time, requiring just 5 epochs compared to the typical 50 epochs needed for classical models.

\subsection{Finance}
The introduction of quantum technology in financial ecosystem tasks has proven useful, especially for critical and complex applications such as fraud detection. QML can play a fundamental role because quantum entanglement can facilitate the capture of intricate relationships, such as those in financial transactions \cite{Grossi_2022}. Additionally, the use of GNNs has proven essential as they can capture and dissect the multifaceted interconnections present in fraud detection. Fraudulent behaviors rarely occur in isolation, so graph-based methods can capture the holistic complexity and unveil concealed connections and dependencies \cite{POURHABIBI2020113303}. Quantum technology plays a crucial role in GNNs, as classical GNNs still have limitations due to their struggle to find complexities in fraud networks and their limitations in handling the explosive growth of data. 

In the work \cite{innan2024financial}, a QGNN is proposed that takes inspiration from the work of \cite{hu2022design}. By comparing tests conducted on the available credit card fraud detection dataset \cite{ULB_MLG_2013} with a classical GNN model, specifically with GraphSAGE \cite{NIPS2017_5dd9db5e}, it was shown that the quantum model outperforms the classical one in terms of performance and accuracy. In fact, despite the simplicity of the quantum model, which consists of only 6 qubits and a single layer with a total of about 200 parameters, the quantum model achieves 94.5\% accuracy, which is better than the 92.4\% accuracy achieved by the classical model. Therefore, QGNNs can represent an important turning point to revolutionize fraud detection and also improve the security of financial systems.

\subsection{Sensor Networks}
Quantum Sensor Networks are a promising area of application for the technologies of Quantum Sensing and Quantum Networking/Communication \cite{kimble2008quantum,qian2019heisenberg}. One useful application is estimating hidden parameters in weak qubit phase rotation signals. For example, this occurs when artificial atoms interact with a small electric field \cite{qian2019heisenberg}. To achieve an advantage in this task, authors in\cite{verdon2019quantum} used a special type of entangled quantum state called a GHZ  state\cite{greenberger1989going}. Surprisingly, even without knowing the entire structure of the quantum network, a QConvGNN can learn to prepare a GHZ state; they used a QConvGNN ansatz with $\hat{H}_1 = \sum_{(j,k) \in E} \sigma_j^z\sigma_k^z$ and $\hat{H}_2 = \sum_{j \in V} \sigma_j^x$. 

The loss function defined to guide the learning process was the negative expectation of the sum of stabilizer group generators (i.e., they ensure that the state being prepared by the QGCNN is a valid GHZ state), which stabilize the GHZ state for a network of $n$ qubits as \cite{toth2005entanglement}:
\begin{equation}
    \mathcal{L}(\boldsymbol{\eta}) = - \langle \bigotimes_{j=0}^n \sigma^x_j + \sum_{j=1}^{n-1}\sigma_j^z\sigma_{j+1}^z \rangle_{\boldsymbol{\eta}}\,.
\end{equation}
Numerical results of the experiments are presented in Fig.~\ref{fig:exp2}, where it is evident that both fidelity and stabilizer expectations reach an excellent plateaus value given enough iterations. The advantage of using such a QConvGNN ansatz for this task is that the number of quantum communication rounds is simply proportional to the number of layers $P$, and that the local dynamics of each node are independent of the global network structure. As a further test of the validity of the GHZ state, the authors also validated the GHZ state by conducting a quantum phase kickback test, which shows how quickly the quantum sensing signal becomes sensitive to changes.


\subsection{Learning Quantum Hamiltonian Dynamics}
As highlighted by \cite{verdon2019quantum,choi2021tutorial}, the high potential of QRecGNN is verified on the time-evolution Ising model Hamiltonian training example.
Learning the dynamics of a closed quantum system is a task of interest for many applications, as discussed in \cite{wiebe2014hamiltonian}. Many researchers \cite{verdon2019quantum,choi2021tutorial} demonstrated that a QRecGNN can learn effective dynamics of an Ising spin system when given access to the output of quantum dynamics at various times. The target is a TIM Hamiltonian on a given graph:
\begin{equation}
\label{eq:qgrnn_formula}
    \hat{H}_\mathrm{target} = \sum_{(j,k) \in E} J_{jk}\sigma^z_j \sigma^z_k + \sum_{v \in V} Q_v \sigma^z_v + \sum_{v \in V} \sigma^x_v\,.
\end{equation}
In terms of quantum gates, $ZZ$ interactions ($\sigma_j^z$ and $\sigma_k^z$ in \eqref{eq:qgrnn_formula}) are represented through parametrized $R_{zz}$ quantum gates; $Z$ terms ($\sigma_v^z$) are represented through parametrized $R_{z}$ quantum gates.

Finally, $X$ terms ($\sigma_v^x$) are represented through simple $R_x$ quantum gates. The initial target TIM Hamiltonian on the near ground-state is obtained using Variational Quantum Eigensolver (VQE) \cite{peruzzoVariationalEigenvalueSolver2014}. They are given copies of a fixed low-energy, non-ground state of the target Hamiltonian $\ket{\psi_0}$, as well as a collection of time-evolved copies of the state $\ket{\psi_t} \equiv \hat{U}(t)\ket{\psi_0} = e^{-it\hat{H}_\mathrm{target}}$ for some known but randomly chosen times $t \in [0,T]$.

The training is performed via the QRecGNN model initialized with an $n$-qubit ($n$-node) complete graph using time-evolution TIM Hamiltonians. The goal is to learn the target Hamiltonian parameters $\boldsymbol{\theta} = \{J_{jk}$, $Q_{v_{j,k,v \in V}}$ by comparing the state $\ket{\psi_t}$ with the state obtained by evolving $\ket{\psi_0}$ according to the QRecGNN ansatz for a number of iterations $P \approx t/\Delta$, where $\Delta$ is a hyperparameter determining the Trotter step size. 

\begin{figure}[!ht]
    \centering
    \includegraphics[width=1.0\columnwidth]{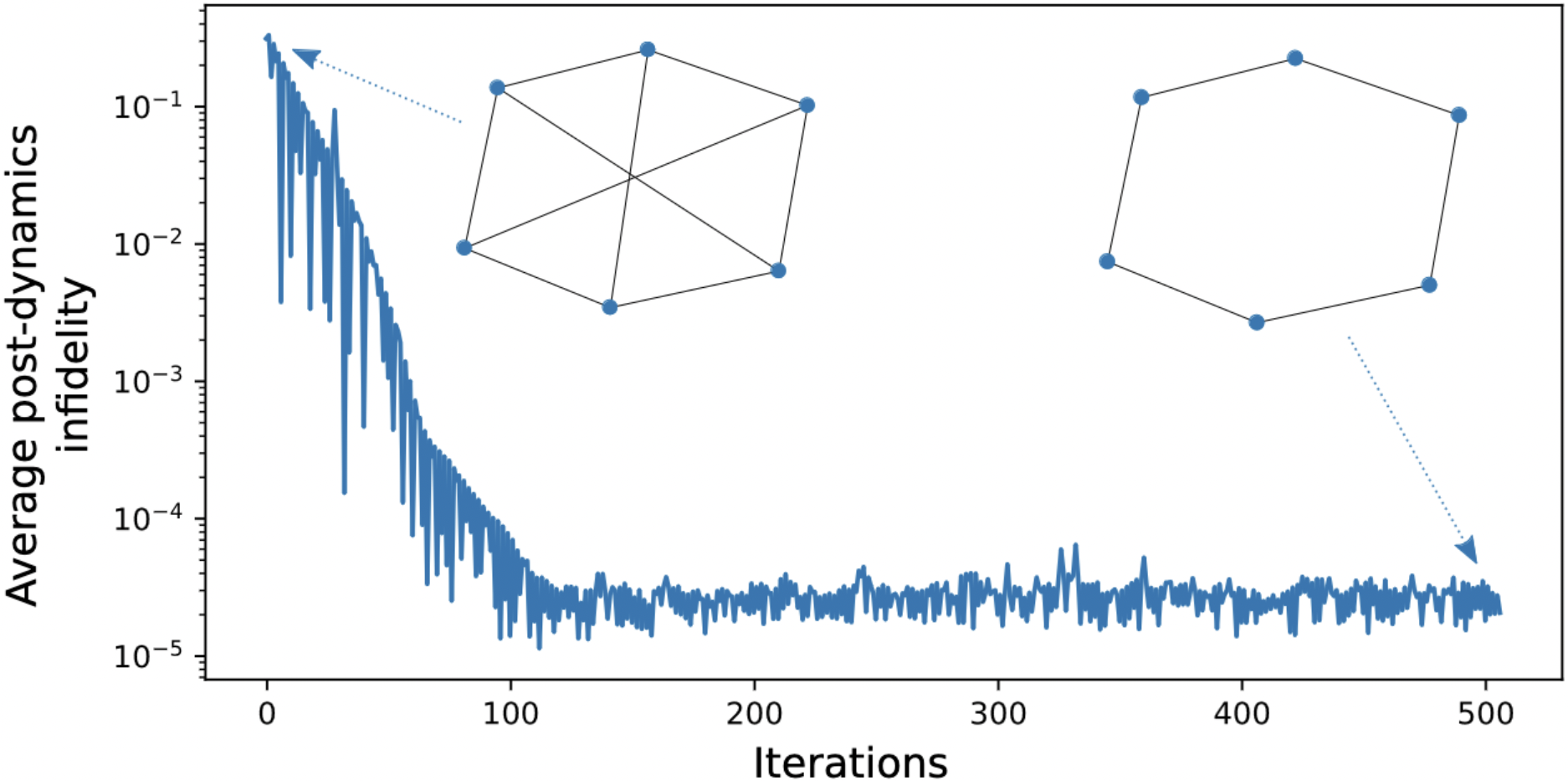}
    \caption{Batch average infidelity with respect to ground truth state sampled at 15 randomly chosen times of quantum Hamiltonian evolution. The initial guess has a densely connected topology and the QRecGNN learns the ring structure of the true Hamiltonian.} 
    \label{fig:learning_original_paper_a}
\end{figure}

\begin{figure}[!ht]
    \centering
    \includegraphics[width=0.6\columnwidth]{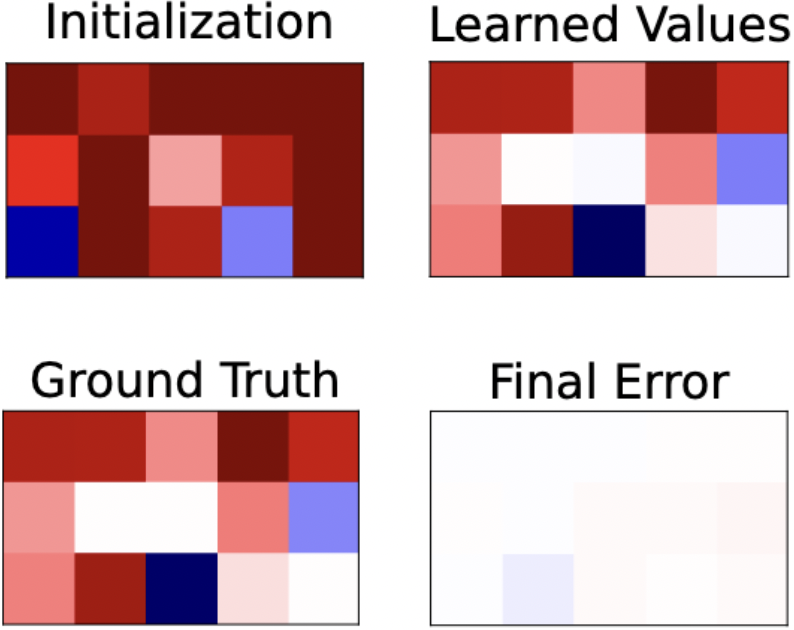}
    \caption{Ising Hamiltonian parameters (weights \& biases) and error on a color scale.} 
    \label{fig:learning_original_paper_b}
\end{figure}

We achieve this by training the parameters via Adam \cite{kingma2014adam} gradient descent on the average infidelity loss function $\mathcal{L}(\boldsymbol{\theta}) = 1 - \frac{1}{B}\sum_{j=1}^B |\bra{\psi_{t_j}| U^j_\mathrm{QGRNN}(\Delta,\boldsymbol{\theta)} \ket{\psi_0}}^2$, where $B$ is the number of quantum data that are used for each step, i.e., batch size. The fidelity $\abs{\bra{\psi_{t_j}} U^j_\mathrm{QGRNN}(\Delta,\boldsymbol{\theta)} \ket{\psi_0}}^2$ measures the quantum state overlap (similarity) between the output of our ansatz $U^j_\mathrm{QGRNN}(\Delta,\boldsymbol{\theta}) \ket{\psi_0}$ and the time-evolved data state $\ket{\psi_{t_j}}$ via the quantum swap test \cite{cincio2018learning}. The ansatz uses a Trotterization of a random densely-connected Ising Hamiltonian with transverse field as its initial guess, and successfully learns the Hamiltonian parameters within a high degree of accuracy as shown in Fig.~\ref{fig:learning_original_paper_a} and Fig.~\ref{fig:learning_original_paper_b}.

The general training process of a QRecGNN for learning a Hamiltonian time-evolution has different features from classical learning methods, e.g., VQE, swap test, and average infidelity loss function. This is because QGRNN handles quantum data. The Ising model emulates Quadratic Unconstrained Binary Optimization (QUBO) problem that is effective in modeling Non-deterministic polynomial (NP) problems, so the potential of QRecGNNs could have great implications for combinatorial optimization as well \cite{papalitsas2019qubo,alom2017quadratic}.

\section{Challenges and Limitations}\label{sec:challenges}
Despite significant advancements in QGNNs, numerous challenges remain that must be overcome to realize the potential for quantum speedups as these architectures are scaled. It is essential to thoroughly characterize the limitations of QGNNs to develop effective strategies for designing more robust algorithms, proving performance guarantees, and ultimately enhancing quantum hardware. Addressing these challenges will be of utmost importance in transitioning from theoretical potential to practical benefits in quantum computing.

\subsection{Noise and Decoherence}
Quantum hardware is inherently susceptible to noise and decoherence, which can significantly degrade the performance of quantum algorithms, including QGNNs. Noise, arising from environmental interactions, imperfect gate operations, and readout errors, can be classified into local (uncorrelated) and correlated types. Decoherence, the loss of quantum information due to these interactions, poses a particular threat to the fragile quantum states used in QGNNs.

Recent studies have investigated the effects of local, uncorrelated noise on VQAs \cite{mccleanTheoryVariationalHybrid2016,mcclean2017hybrid,gentini2020noise,fontana2021evaluating}. For example, authors in \cite{kungurtsev2022iteration} emphasize that the inherent noise in near-term quantum devices introduces systematic bias in objective function evaluations. This bias necessitates a reevaluation of convergence analysis for classical optimization procedures in quantum contexts. The quantum system's susceptibility to various local noise channels, such as dephasing, bit-flip, and depolarizing, further complicates this analysis. Correlated errors, including crosstalk \cite{heinz2021crosstalk}, non-Markovian $1 / f$ noise \cite{burkard2009non}, and interactions with environmental fluctuators \cite{schlor2019correlating}, have been shown to be prevalent in NISQ devices, further complicating the implementation of QGNNs. 

While quantum error correction (QEC) codes can mitigate hardware noise, they require significant qubit and gate operation overhead \cite{roffe2019quantum}. Quantum error mitigation techniques such as zero-noise extrapolation \cite{temme2017error}, probabilistic error cancellation \cite{van2023probabilistic} and their combination \cite{mari2021extending} have shown promise on NISQ devices, but face scalability and large-scale implementation challenges.

A significant gap in current research is the absence of comprehensive theoretical and empirical studies that characterize the performance degradation of QGNN as a function of noise strength and graph size. As quantum noise increases, we expect QGNN performance to degrade. However, the precise nature of this degradation - whether it's linear, exponential, or follows some other pattern - remains unclear. The complexity of the input graph likely influences how severely noise affects QGNN performance: larger graphs may be more susceptible to noise-induced errors, as they require longer circuits. Moreover, deeper circuits with more layers potentially offer better expressibility and graph representation capabilities, but they also provide more opportunities for errors to accumulate due to prolonged exposure to noise. 

If we had a quantitative understanding of these relationships, it would enable us to make informed design decisions about QGNN architecture for real-world applications, could facilitate the determination of optimal circuit depth for specific noise levels and graph sizes, and lead to the development of adaptive QGNN designs that dynamically adjust their structure based on input graph properties and quantum hardware noise characteristics. One reason for this research gap may be the current limitations in simulating and executing large QGNNs on actual quantum hardware, as illustrated in the following Sect.~\ref{subsec:scalability}. 

A primary strategy for noise mitigation is reducing gate count in compiled quantum circuits, as different hardware architectures impact circuit depth and connectivity. In this sense, developing QGNN-specific graph-to-circuit mapping and noise-resilient techniques that exploit graph data structure could be a promising research direction. This approach could leverage graph connectivity and redundancy for more efficient error correction strategies, potentially leading to feasible QGNN implementations on NISQ devices. 

\subsection{Scalability and Adaptability}
\label{subsec:scalability}
Scalability is a critical challenge for QGNNs. As the size of the graph and the number of qubits increase, the complexity of the quantum circuits also grows, leading to greater demands on quantum hardware. Current QGNN approaches face the challenge of qubit demand growing linearly with graph size, making large-scale graph analysis on NISQ devices impractical. 

One approach to addressing scalability is using topology aware quantum circuit synthesis for qubit-efficient encoding and mappings of graph data onto quantum circuits \cite{davis2020towards,weiden2022wide}. Unitary synthesis techniques offer optimal gate counts while respecting hardware-specific qubit topologies. Moreover, circuit optimization strategies, including gate pruning, reinforcement learning and parallelization, can help manage circuit depth and execution times \cite{nam2018automated,ge2024quantum,sim2021adaptive,ostaszewski2021reinforcement}.

Hybrid quantum-classical approaches also offer a practical solution for scalability. By offloading parts of the computation to classical layers, it is possible to handle larger graphs that would be otherwise unfeasible to process on quantum hardware alone. This approach has been successfully applied in \cite{chen2021hybrid,mauro2024hybrid,tuysuz2021hybrid}.

\subsection{Lack of Performance Guarantees}
Unlike classical ML and QAOA-based algorithms, where performance can often be analyzed and bounded theoretically, QGNNs currently lack robust performance guarantees. This gap arises from the inherent challenges in theoretical analysis and the necessity for empirical validation to evaluate performance effectively. The diverse, fragmented and heuristic nature of QGNN architectures (see Sect.~\ref{sec:qgnns}) further complicate this issue.

Recent works on VQAs have begun to partially address this gap. For example, several studies have provided bounds on the expressiveness of QNNs, suggesting conditions under which quantum models might outperform classical ones \cite{du2020expressive, abbas2021power,yang2024maximizing}. However, these results are preliminary and often depend on idealized assumptions about ansatz design, quantum hardware and noise levels. 

A promising research direction is the development of theoretical frameworks that can provide more concrete performance guarantees for QGNNs. This includes understanding the role of entanglement and quantum correlations in enhancing the expressive power of QGNNs. Additionally, benchmarking QGNNs against classical counterparts on a variety of graph-based tasks could provide empirical insights into their practical advantages and limitations.

\subsection{Barren Plateaus}
Training variational QNNs, including QGNNs, is often significantly challenged by the barren plateau phenomenon \cite{mcclean2018barren}. Unlike classical NNs, where gradients vanish as the number of layers increases, QNNs encounter this issue more broadly as the number of qubits grows. Despite \cite{Vitz2024HybridQG} used a gradient-free optimizer for their QGNN, barren plateaus may hinder the training of quantum models designed for high-dimensional data, even when using gradient-free optimization algorithms \cite{arrasmith2021effect}. The barren plateau problem is further exacerbated by quantum noise \cite{wang2021noise} and the design of the cost function \cite{cerezo2021cost}. Notably, there has been experimental evaluation of local cost functions compared to global cost functions in addressing the barren plateau issue within QGNNs \cite{ryu2023quantum}. One potential reason for the observed improvement is that optimization algorithms tend to encounter barren plateaus more frequently with global readout functions, as suggested by \cite{cerezo2021cost}. However, the difference in performance is minimal, and \cite{ryu2023quantum} provides limited insight into barren plateaus for QGNNs.

Moreover, the architecture of the quantum circuit plays a crucial role in the presence of barren plateaus. For instance, EQGCs proposed in \cite{mernyei2023equivariant} have been observed to scale effectively with model depth and are less prone to barren plateaus than other QNNs architecture. Nonetheless, these findings are based on circuits with only 6 to 10 qubits, while barren plateaus are more likely to emerge in deeper circuits, necessitating further empirical validation. Overall, while the barren plateau problem in QGNNs is widely recognized, it remains insufficiently addressed in the literature. Several training strategies and initialization techniques have been proposed to mitigate this issue \cite{grant2019initialization,zhang2022escaping} (see Sect.~\ref{subsec:initialization}), but further research and development are required to fully understand and resolve the challenges posed by barren plateaus in variational quantum graph-based models.

\subsection{Initialization Strategies}
\label{subsec:initialization}
Quantum circuits' parameters initialization in QGNNs is a critical step that can significantly impact their performance. Random initialization can lead to barren plateaus \cite{mcclean2018barren}, therefore effective initialization strategies are needed to provide good starting points for the parameters of quantum circuits.

Despite its importance, the topic of initialization is not extensively covered in the literature, and many papers give it minimal attention. For example, \cite{ryu2023quantum} compared EDU-QGC with classical Graph Neural Networks, using three different training runs with varied initial parameters for the classical models to find optimal weights. However, for the quantum and hybrid models, only a single training run was conducted, with all initial quantum weights set to 1. Such an evaluation approach is puzzling, given that the classical models' training results showed a significant dependence on initial weight values. The decision to use a single, uniform initialization for quantum models might be influenced by performance considerations or limitations specific to quantum systems, but it calls for further investigation and clarity. Additional evidence of the importance of initialization strategies is found in \cite{mernyei2023equivariant}, where a EDU-QGC model experienced a failure due to bad initialization, resulting in a poor 50\% accuracy. Despite using only a small number of qubits (6 to 10), unfortunate starting points can lead to significant learning issues.

Heuristic-based initialization methods, such as initializing parameters based on classical solutions or using domain-specific knowledge, have been shown to improve convergence in VQAs \cite{grant2019initialization}. Another approach is to use warm-start techniques, where the parameters are initialized close to known good solutions from related problems or smaller instances of the current problem \cite{truger2024warm}.
Techniques such as Gaussian initialization \cite{zhang2022escaping}, which uses Gaussian distributions to set initial parameters, and layer-wise training \cite{skolik2021layerwise}, where parameters are initialized progressively from shallow to deeper layers, could provide promising alternatives. These approaches might help mitigate the barren plateau problem by ensuring more favorable starting conditions for the optimization process.

\section{Conclusion}\label{sec:conclusions}
QGNNs are still an unexplored yet promising research field, as demonstrated by the above papers and preliminary experiments. Experimental findings indicate that QGNNs are able to efficiently learn quantum dynamics by encoding the problem Hamiltonian into quantum circuits or performing prediction tasks on graphs using few qubits and less parameters than the classical counterparts, thus enabling us to create quantum graph neural models with superior performances.  

The analysis and outcomes presented in this study should not be regarded as definitive conclusions about the QGNN approach, nor should the reader be discouraged by current limitations. Instead, this comprehensive analysis and review work should be seen as an encouraging initial foray into exploring the potential applications of QGNNs. Through several numerical experiments presented in literature, we have shown that the use of different QGNN ansatzes may provide promising results in the context of quantum dynamics learning, quantum sensor network optimization,
unsupervised graph clustering, supervised graph isomorphism classification, image classification, particle track reconstruction, and many more graph-oriented problems. 

A common pattern that emerges from the study of the literature is that the QGNN model proposed by \cite{verdon2019quantum} and its variants are very similar to QAOA \cite{farhi2014quantum}, in the sense that they encode the structure of the problem (graph) directly into the quantum circuit, thus not relying on any classical layer. Many QAOA variants have been proposed in the literature \cite{blekos2023review}, so one could wonder if those enhancement may also apply to QGNNs. In addition, some QAOA's interesting properties could be investigated in the QGNNs realm as well, such as symmetry in the model's parameters and the relation between graph's properties and QGNNs' performance.

In a subsequent investigation, a challenge which has not been addressed yet relates to QGNN's performance under noise and lack of quantum resources: as the QAOA, QGNNs might be severely impacted by noise in NISQ devices, as well as by qubit scarcity, and thus become useless for very large graphs on modern quantum computers. Furthermore, the topology of quantum hardware (i.e., qubits connectivity) should also be taken into careful consideration when designing QGNN circuits. 
Another hot topic which could be worth exploring concerns the trainability of such QGNN models and their possible resilience to barren plateaus, which are now afflicting the QNN landscape.

Considering the extensive body of literature concerning the application of GNNs and their derivatives in quantum chemistry, forthcoming studies should investigate quantum approaches to molecule simulations, where one can learn a graph-based hidden quantum representation of a quantum chemical process via a QGNN. Given that the genuine underlying process is fundamentally quantum and possesses an inherent molecular graph structure, the QGNN has the potential to serve as a more precise model for the latent processes that give rise to observed emergent chemical characteristics. 

On the other hand, hybrid quantum-classical models as the ones proposed by \cite{chen2021hybrid,tuysuz2021hybrid} might leverage VQC's properties like high-dimensional quantum feature map, entanglement and data processing in a complex state space to enhance the performance of the classical GNN model without worrying about the number of qubits and the graph size. Other future work could include generalizing the QGNN to include quantum degrees of freedom on the edges and extending the QGNN architecture to handle and model hypergraphs and simplicial complexes.

\section*{Acknowledgments}
The contribution of A. Ceschini, F. De Falco, A. Rosato and M. Panella in this work was in part supported by the ``NATIONAL CENTRE FOR HPC, BIG DATA AND QUANTUM COMPUTING'' (CN1, Spoke 10) within the Italian ``Piano Nazionale di Ripresa e Resilienza (PNRR)'', Mission 4 Component 2 Investment 1.4 funded by the European Union - {NextGenerationEU} - CN00000013 - CUP B83C22002940006.

The contribution of A. Verdone in this work was in part supported by the Italian Ministry of University and Research (MUR), which funded his PhD grant in Information and Communication Technology (ICT) as per the Ministerial Decree no. 1061/2021. 

\bibliographystyle{IEEEtran}
\bibliography{IEEEabrv,main}

\end{document}